\documentclass[twocolumn,amsmath,amssymb,aps,prd,10pt,superscriptaddress,nofootinbib,noeprint,preprintnumbers,floatfix,notitlepage]{revtex4-2}

\pdfoutput=1

\usepackage{graphicx}
\usepackage{dcolumn}
\usepackage{bm}
\usepackage{mathtools}
\usepackage{multirow}
\usepackage{amssymb}
\usepackage{amsmath}
\usepackage{amsfonts}
\usepackage{xcolor}
\usepackage{tabularx}
\usepackage{array}
\usepackage[utf8]{inputenc}
\usepackage{hyperref}
\usepackage[letterspace=-10]{microtype}
\usepackage{soul}

\definecolor{jlab_red}{RGB}{192,39,45}
\definecolor{jlab_orange}{RGB}{249,102,0}
\definecolor{jlab_blue}{RGB}{47,122,121}
\definecolor{jlab_green}{RGB}{65,125,10}

\newcommand{\DDst}{\ensuremath{{DD^\ast}}}
\newcommand{\DstDst}{\ensuremath{{D^\ast{D}^\ast}}}

\newcommand{\SLJ}[3]{\ensuremath{{\:\!}^{{#1}\!}{#2}_{#3}}}
\newcommand{\SLJc}[3]{\ensuremath{  \{  {\:\!}^{{#1}\!}{#2}_{#3}  \}  }}
\newcommand{\cm}{{cm}}

\hypersetup{%
pdftitle = {Resonances in coupled DD*-D*D* scattering from QCD},
pdfsubject = {QCD},
pdfkeywords = {QCD, Hadron, Charm, Lattice, Meson, Scattering},
pdfauthor = {Hadron Spectrum Collaboration},
colorlinks = {true},
filecolor = {black},
linkcolor = {jlab_blue},
menucolor = {black},
citecolor = {jlab_green},
urlcolor = {jlab_green},
}{}

\begin{document}

\preprint{arXiv:2405.15741}

\title{Near-threshold states in coupled $DD^{\ast}-D^{\ast}D^{\ast}$ scattering from lattice QCD}

\author{Travis Whyte}
 \email{t.whyte@fz-juelich.de}
\affiliation{
Department of Applied Mathematics and Theoretical Physics, Center for Mathematical Sciences,\\University of Cambridge, Wilberforce Road, Cambridge, CB3 0WA, UK
}
\affiliation{
School of Mathematics, Trinity College, Dublin 2, Ireland
}
\author{David J. Wilson}
\email{d.j.wilson@damtp.cam.ac.uk}
\affiliation{
Department of Applied Mathematics and Theoretical Physics, Center for Mathematical Sciences,\\University of Cambridge, Wilberforce Road, Cambridge, CB3 0WA, UK
}
\author{Christopher E. Thomas}
\email{c.e.thomas@damtp.cam.ac.uk}
\affiliation{
Department of Applied Mathematics and Theoretical Physics, Center for Mathematical Sciences,\\University of Cambridge, Wilberforce Road, Cambridge, CB3 0WA, UK
}
\collaboration{for the Hadron Spectrum Collaboration}
\date{Feb. 24th, 2025}

\begin{abstract}
\noindent
The first determination of doubly-charmed isospin-0 coupled-channel $DD^\ast-D^\ast D^\ast$ scattering amplitudes from lattice QCD is presented. The finite-volume spectrum is computed for three lattice volumes with a light-quark mass corresponding to $m_\pi\approx 391$ MeV and is used to extract the scattering amplitudes in $J^P = 1^+$ via the L\"{u}scher quantization condition. By analytically continuing the scattering amplitudes to complex energies, a $T_{cc}$ pole corresponding to a virtual bound state is found below $DD^\ast$ threshold. We also find a second pole, $T_{cc}^\prime$, corresponding to a resonance pole below the kinematically closed $D^\ast D^\ast$ channel, to which it has a strong coupling. A non-zero coupling is robustly found between the $S$-wave $D D^\ast$ and $D^\ast D^\ast$ channels producing a clear cusp in the $D D^\ast$ amplitude at the $D^\ast D^\ast$ threshold energy. This suggests that the experimental $T_{cc}^\prime$ should be observable in $D D^\ast$ and $D^\ast D^\ast$ final states at ongoing experiments.
\end{abstract}

\maketitle

\section{\label{sec:intro}Introduction}

The wealth of exotic hadron states discovered recently in experimentation has driven theoretical efforts to understand their nature. In particular, the recent observation of the doubly-charmed, manifestly exotic $T_{cc}^+(3875)$ at the LHCb experiment \cite{LHCb:2021vvq,LHCb:2021auc} close in energy to $\DDst$ threshold motivates further study. Having two charm quarks and being charged, this state goes beyond expectations from conventional quark model $q\bar{q}$ spectra, with a minimal $cc\bar{u}\bar{d}$ quark content content~\cite{LHCb:2021vvq}. LHCb investigates several charged and neutral $\DDst$ combinations, finding only one enhancement and implying a preference for isospin $I=0$.

Several aspects of the $T_{cc}$ make it particularly appealing for theoretical studies. The exotic nature limits the possible decay modes, it arises close to $\DDst$ threshold and it is very narrow. This apparent simplicity should help in determining the relevant degrees of freedom. However, the $T_{cc}$ and the $D^\ast$ through which it decays are unstable resonances and undergo decays to lighter hadrons.
The unstable nature of the $D^\ast$ leads to a three-body $DD\pi$ final state. Furthermore, two channels open close together, due to the different charge combinations -- the energy splitting of the $D^0D^{\ast+}$ and $D^+D^{\ast0}$ thresholds is of the order of the observed width and so could be significant.

The intrinsic spin of the vector $D^\ast$ can combine with the $D$-meson in many ways. Non-zero angular momenta are suppressed close to threshold and so in the theoretical modeling of this system, it has been widely assumed that $S$-wave is dominant, forming $J^P=1^+$. However, higher angular momenta should also be investigated.

The LHCb $\DDst$ data spans a 30 MeV window up to 3900 MeV~\cite{LHCb:2021vvq,LHCb:2021auc}. A little higher in energy the $\DstDst$ channel opens, which may bring additional features.
A variety of theoretical models find a state with quantum numbers $J^P = 1^+$ below $\DstDst$ threshold \cite{Molina:2010tx,Li:2012ss,Sun:2012zzd,Liu:2019stu,Dong:2021bvy,Dai:2021vgf,Ortega:2022efc}. Some have associated this state as a heavy partner of the $T_{cc}$ seen near $\DDst$ threshold. A prediction of a partner to the $T_{cc}$ based on fits to the LHCb data and Heavy Quark Spin Symmetry \cite{Neubert:1993mb} was given in \cite{Albaladejo:2021vln}. Subsequent predictions from chiral effective field theory, including three body effects but assuming $\DDst$ and $\DstDst$ are decoupled, appear in \cite{Du:2021zzh}. Its properties and decay modes were studied in \cite{Jia:2022qwr,Jia:2023hvc}.

Quantum Chromodynamics (QCD) is the theory of the strong interactions and it is thus desirable to know what QCD predicts for the $T_{cc}$. A nonperturbative approach to QCD making only controlled and systematically improvable approximations is \emph{lattice QCD}, where the fundamental equations of interacting quarks and gluons are discretized on a finite spacetime lattice. The QCD path integral is sampled on the lattice using Monte Carlo techniques, which allows for the extraction of correlation functions and the calculation of the spectrum of the eigenstates of QCD on the lattice. Lattice calculations are performed in a finite spatial volume and we use periodic spatial boundary conditions. As such, the momentum is quantized and a discrete spectrum is obtained.

The discrete spectra can then be used to determine infinite volume scattering amplitudes that are continuous functions of energy using the L\"{u}scher finite volume formalism \cite{Luscher:1990ux,Rummukainen:1995vs,Bedaque:2004kc,Kim:2005gf,He:2005ey,Lage:2009zv,Fu:2011xz,Leskovec:2012gb,Gockeler:2012yj,Hansen:2012tf,Briceno:2012yi,Guo:2012hv,Briceno:2014oea}. The scattering amplitudes contain singularities, which are a model independent way of understanding the spectroscopic content in terms of bound states, virtual bound states, and resonances. The approach is reviewed in Ref.~\cite{Briceno:2017max}.

Lattice calculations of resonances typically use degenerate light-quark masses that produce a pion that is too massive. When considering charm quarks, this also results in a $D^\ast$ that is stable against decay to $D\pi$ for all but the very smallest light-quark masses. This enables a simple treatment of $\DDst$ scattering using well-established two-hadron methods which have demonstrated success in charmed~\cite{Mohler:2012na,Mohler:2013rwa,Moir:2016srx,Gayer:2021xzv,Cheung:2020mql} and charmonium systems~\cite{Lang:2014yfa,Lang:2015sba,Prelovsek:2020eiw,Wilson:2023hzu,Wilson:2023anv}.

Investigations of $\DDst$ on the lattice have been made over a range of light quark masses. Ref. \cite{Junnarkar:2018twb} found that the ground state energy obtained utilizing a basis of meson-meson and tetraquark operators in $J^P = 1^+$ lies below the $\DDst$ threshold using three lattice spacings and $m_{\pi}$ ranging from 153 to 689 MeV. This is suggestive of the presence of attractive interactions, however the existence of a pole was not verified. In Ref. \cite{Lyu:2023xro}, the HAL QCD method was used to extract $S$-wave phase shifts with pions at nearly physical mass, locating a virtual bound state 45 keV below $\DDst$ threshold. Ref. \cite{Chen:2022vpo} examined elastic $\DDst$ scattering in $S$-wave for $I=0,1$, but did not verify the presence of a pole. 
Elastic $\DDst$ $S$-wave scattering was investigated in \cite{Padmanath:2022cvl} for $m_{\pi} \approx 280$ MeV, and a pole in the scattering amplitudes corresponding to a virtual bound state was found $9_{-7}^{+4}$ MeV below the $\DDst$ threshold. Its quark mass dependence and the role of the left hand cut due to one pion exchange between $D$ and $D^\ast$ were examined in \cite{Collins:2024sfi}. A left hand cut due to one pion exchange between $D$ and $D^\ast$ is also present in this calculation, as well as a second left hand cut due to one pion exchange between $D^\ast$ and $D^\ast$ in the coupled-channel case. 

In this article, the first calculation of coupled-channel isospin-0 $J^P = 1^+$ $\DDst-\DstDst$ scattering from lattice QCD is presented, finding a virtual bound-state below $\DDst$ threshold and a \emph{resonance} just below $\DstDst$ threshold coupled to both channels. In Section \ref{sec:symm}, the relevant partial wave configurations and symmetries for $\DDst$ and $\DstDst$ are discussed. In Section \ref{sec:lattice_energies}, the techniques used to extract the finite volume spectra are reviewed, including operator construction and variational analysis of the computed correlation functions. The finite volume spectra obtained from both systems overall at rest and with non-zero total momentum for three lattice volumes is also presented. In Section \ref{sec:scattering}, the resulting scattering amplitudes extracted from the finite volume spectra using the L\"{u}scher formalism are presented, including the scattering amplitudes of higher partial waves. Section \ref{sec:poles} contains an investigation of the singularities of the amplitudes. An interpretation is given in Section \ref{sec:interp} and we summarize our results in Section \ref{sec:summary}.

\section{\label{sec:symm}Partial waves}

We first consider the possible partial waves and their relevant symmetries for $\DDst$ and $\DstDst$ with exact isospin symmetry. These are classified by their total spin parity $J^P$, but for any given $J^P$ different orbital angular momenta may contribute due to the presence of the vector $D^\ast$. It is convenient to use a spectroscopic notation $\SLJ{2S+1}{\ell}{J}$ with total spin $S$ and orbital angular momentum $\ell$. Partial waves grow at threshold like $k_\cm^{2\ell}$ in the absence of any enhancement, where $k_\cm = \sqrt{(s-(m_{D^{(\ast)}} + m_{D^{\ast}})^2)(s-(m_{D^{(\ast)}} - m_{D^{\ast}})^2)/4s}$ is the scattering momentum in the center-of-mass (cm) frame and $s = E_\cm^2$ with $E_\cm$ the total energy in the cm-frame.
 
In $J^P=1^+$, the leading $\DDst$ contribution is in $S$-wave, but it may also mix dynamically with $D$-wave to form a coupled $\SLJ{3}{S}{1}-\SLJ{3}{D}{1}$ system. $P$-waves produce $J^P=0^-,1^-,2^-$ combinations, and further $D$-wave combinations arise in other $J^P$, such as $3^+$.
The relevant combinations up to $F$-wave are summarized in Table~\ref{tab:DDstPWs}.

Due to the presence of identical bosons in $\DstDst$ states and the requirement that the total wavefunction be symmetric under the interchange of the two $D^{\ast}$, many $\DstDst$ combinations are forbidden. In $I=0$ the flavor wavefunction is antisymmetric, so the spin$\times$space wavefunction is also required to be antisymmetric. This leads to the nonzero partial waves given in Table \ref{tab:DstDstPWs}, with $S=1$ for $\ell$-even and $S=0,2$ for $\ell$-odd.

\begin{table}
\center
\begin{ruledtabular}
\caption{\label{tab:DDstPWs} The allowed partial wave configurations up to $\ell = 3$ for $\DDst$ in $I=0$.}
\begin{tabularx}{0.66\columnwidth}{ccc}
&&\\[-2.5ex]
$\ell$ & $\SLJ{2S+1}{\ell}{J}$ & $J^P$ \\
\hline
\hline
&&\\[-2.5ex]
	0 & \SLJ{3}{S}{1} & $1^+$ \\
\hline
&&\\[-2.5ex]
	1 & \SLJ{3}{P}{0,1,2} & $\{0,1,2\}^-$ \\
\hline
&&\\[-2.5ex]
	2 & \SLJ{3}{D}{1,2,3} & $\{1,2,3\}^+$ \\
\hline
&&\\[-2.5ex]
	3 & \SLJ{3}{F}{2,3,4} & $\{2,3,4\}^-$ \\
	\end{tabularx}
\end{ruledtabular}
\end{table}

\begin{table}
\center
\begin{ruledtabular}
\caption{\label{tab:DstDstPWs} The allowed $\DstDst$ $I=0$ partial wave configurations up to $\ell = 3$ and the symmetry (symmetric ``S" or antisymmetric ``A") for the spin, space, and flavor  wavefunctions which result in an overall wavefunction that is symmetric under the interchange of the two $D^\ast$.}
\begin{tabular}{ccccccc}
	&&&&&\\[-2.5ex]
	$\ell$ & $\SLJ{2S+1}{\ell}{J}$ & $J^P$ & Spin & Space & Flavor & Total\\
\hline
\hline
&&&&&\\[-2.5ex]
	0 & \SLJ{3}{S}{1} & $1^+$ & A & S & A & S \\
\hline
&&&&&\\[-2.5ex]
	\multirow{2}{*}{1} & \SLJ{1}{P}{1} & $1^-$ & \multirow{2}{*}{S} & \multirow{2}{*}{A} & \multirow{2}{*}{A} & \multirow{2}{*}{S} \\
	& \SLJ{5}{P}{1,2,3} & $\{1,2,3\}^-$  &  &  &  \\
\hline
&&&&&\\[-2.5ex]
	2 & \SLJ{3}{D}{1,2,3} & $\{1,2,3\}^+$ & A & S & A & S \\
\hline
&&&&&\\[-2.5ex]
	\multirow{2}{*}{3} & \SLJ{1}{F}{3} & $3^-$ & \multirow{2}{*}{S} & \multirow{2}{*}{A} & \multirow{2}{*}{A} & \multirow{2}{*}{S} \\
	& \SLJ{5}{F}{1,...,5} & $\{1,2,3,4,5\}^-$ &  &  &  &  \\
\end{tabular}

\end{ruledtabular}
\end{table}

\section{\label{sec:lattice_energies}Determination of Lattice Energies}
 
\subsection{\label{sec:lattice_setup}Lattice Setup}

Lattice calculations were performed on anisotropic lattices with three volumes having the same quark mass and lattice spacing. The volumes and the number of configurations per volume are given in Table~\ref{tab:latt_params}. In order to improve energy resolution, the temporal lattice spacing $a_t$ is chosen to be finer than the spatial one $a_s$ by a factor $\xi  = a_s / a_t \approx 3.5$. Gauge configurations were generated using a tree level improved Symanzik action in the gauge sector and a Wilson-Clover action with $2+1$ flavors of dynamical quarks in the fermion sector, where the strange quark has been tuned to approximately its physical value \cite{Edwards:2008ja,Lin:2008pr}. The mass of the $\Omega$ baryon is used to set the scale when calculating quantities in physical units, which is $a_t^{-1} = m^{\Omega}_{phys} / a_t m_{\Omega} = 5667$ MeV \cite{Edwards:2012fx}, giving a pion mass of $m_{\pi} \approx 391$ MeV. This corresponds to a spatial lattice spacing $a_s\approx 0.12$~fm. The quenched charm quark also uses the Wilson-Clover action and its mass is tuned to approximately reproduce the physical mass of the $\eta_c$ meson as discussed in \cite{Liu:2012ze}, which results in a bare mass of $a_tm_c = 0.092$. This results in stable $D$ and $D^{\ast}$ mesons with masses approximately $1886$ MeV and $2010$ MeV respectively.  See Table~\ref{tab:masses} for the masses of the relevant stable mesons in this study. In the scattering analyses, we use the anisotropy determined from the dispersion relation of the pion as the central value, which was found to be $\xi = \xi_{\pi} = 3.444(6)$ \cite{Dudek:2012gj}. However, to account for the systematic uncertainty from obtaining different anisotropies from the relativistic dispersion relations for the $\pi$, $D$ and $D^\ast$ mesons across the three lattice volumes, we utilize a value of $\xi = 3.444(50)$ in this analysis. This uncertainty spans the central values of the anisotropies determined from the $\pi$, $D$ and $D^\ast$ mesons across the three lattice volumes and the uncertainty is propagated through to the cm-frame energies obtained from the lattice energies. The lattice energies are calculated in the lattice frame where the $D^{(\ast)}D^\ast$ system can have overall nonzero momentum and are translated to the cm-frame where it has overall zero momentum. Consequently, this is also propagated through into the scattering amplitudes.  See \cite{Cheung:2020mql,Wilson:2023anv} for further details of the dispersion relations.

\begin{table}
\center
\begin{tabular}{c|c|c|c}
$(L/a_s)^3 \times (T/a_t)$ & $N_\mathrm{cfgs}$ & $N_\mathrm{tsrcs}$ & $N_\mathrm{vecs}$ \\
\hline
& & &\\ [-2.25ex]
$16^3 \times 128$ & 478 & 8 & 64  \\
$20^3 \times 256$ & 288  & 3-4 & 128 \\
$24^3 \times 128$ & 553 & 3-4 & 160 \\
\hline
\end{tabular}
	\caption{\label{tab:latt_params}A summary of the details for each ensemble used in this work. $(L/a_s)^3 \times (T/a_t)$ denotes the volume of the lattice with $L$ the spatial extent and $T$ the temporal extent of the lattice. $N_\mathrm{cfgs}$ denotes the number of gauge-field configurations used. $N_\mathrm{tsrcs}$ denotes the number of time sources used per gauge-field configuration. On the $(L/a_s)^3 = 20^3,24^3$ volumes, four time sources were used to calculate the spectrum for overall zero momentum and three for overall nonzero momentum.} $N_\mathrm{vecs}$ denotes the number of vectors used in the distillation basis.
\end{table}

\begin{table}
\begin{tabular}{ccc}
&&\\
\end{tabular}
\begin{tabular}{c|c}
\multicolumn{2}{c}{}\\
& $a_t m$ \\
\hline
$\pi$ & 0.06906(13)\\
$D$ & 0.33281(9)\\
$D^{\ast}$ & 0.35464(14)\\
$D^{\ast}_0$ & 0.40170(18) \\
\hline
\end{tabular}
\begin{tabular}{ccc}
&&\\
\end{tabular}
\begin{tabular}{c|c}
\multicolumn{2}{c}{}\\
& $a_t m$  \\
\hline
$DD^{\ast}$ & 0.68745(17)\\
$D^{\ast}D^{\ast}$ & 0.70928(20)\\
$DD^{\ast}_0$ & 0.73381(20) \\
$DD\pi$ & 0.73468(18) \\
\hline
\end{tabular}
	\caption{\label{tab:masses}(Left) The relevant masses for the stable $\pi$ \cite{Dudek:2012gj}, $D$, $D^{\ast}$ \cite{Wilson:2023anv} and $D_0^{\ast}$ \cite{Moir:2016srx} and (right) kinematic thresholds for this study.}
\end{table}

\subsection{\label{sec:ops}Details of the Calculation}
In order to extract the finite volume spectra from correlation matrices, we follow the well-established variational method of the generalized eigenvalue problem (GEVP) \cite{Blossier:2009kd,Michael:1985ne,Luscher:1990ck} and summarize our implementation \cite{Dudek:2010wm} here. Given a basis of operators $\{\mathcal{O}_i\}$ that interpolate the quantum numbers of interest, we solve the GEVP $C_{ij}(t) v^{\mathfrak{n}}_j(t)= \lambda_{\mathfrak{n}}(t,t_0) C_{ij}(t_0) v^{\mathfrak{n}}_j(t)$ where
\begin{equation}
C_{ij}(t) = \langle 0 \vert \mathcal{O}_i(t) \mathcal{O}_j^\dagger(0) \vert 0 \rangle%
\label{eq:corr_matrix}
\end{equation}
is the matrix of correlation functions and $t_0$ is an appropriately chosen reference timeslice. The eigenvalue, or ``principal correlator'', $\lambda_{\mathfrak{n}}(t,t_0)$, is used to obtain the energy $E_{\mathfrak{n}}$ of the $\mathfrak{n}^{th}$ eigenstate $\vert \mathfrak{n} \rangle$ through a fit to the form $\lambda_{\mathfrak{n}}(t,t_0) = (1-A_{\mathfrak{n}})e^{-E_{\mathfrak{n}}(t-t_0)} + A_{\mathfrak{n}} e^{-E_{\mathfrak{n}}^{\prime}(t-t_0)}$, where the second exponential is used to allow for contamination from excited states. The eigenvectors, $v^{\mathfrak{n}}_j(t)$, give the operator-state overlaps 
\begin{equation}
	Z_i^{\mathfrak{n}} = \langle \mathfrak{n} \vert \mathcal{O}_i^\dagger(0) \vert 0 \rangle = \sqrt{2E_{\mathfrak{n}}}e^{(E_{\mathfrak{n}}t_0)/2}(v^{\mathfrak{n}}_j)^*C_{ji}(t_0) \, ,
\end{equation}
and are used to construct variationally optimized operators $\Omega_{\mathfrak{n}}^{\dagger} = \sum_i v^{\mathfrak{n}}_i \mathcal{O}_i^{\dagger}$, the optimal linear combination of the basis of operators to interpolate state $\mathfrak{n}$.

Because the finite spatial volume and discretization of the lattice break the rotational symmetry of the infinite volume continuum, angular momentum $J$ is no longer a good quantum number, and mesons at rest with definite continuum $J$ are labelled by the finite cubic group irreducible representation (irrep) $\Lambda$, which we refer to as \emph{subduction} \cite{Johnson:1982yq,Dudek:2010wm}. Additionally, the rotational symmetry is further broken for mesons in flight, and so such states are labelled by an irrep of the little group LG$(\vec{P})$ \cite{Moore:2005dw,Thomas:2011rh}. The subductions for pseduoscalar-vector and vector-vector scattering relevant to $D^{(\ast)}D^{\ast}$ can be found in Appendix \ref{sec:sub_tables}. 

The basis of interpolating operators $\{\mathcal{O}_i\}$ must be chosen to be sufficiently large and with suitable structures such that all levels in the energy region of interest can be extracted reliably. We employ meson-meson-like operators \cite{Dudek:2012gj} of the form
\begin{equation}
\begin{split}
	\mathcal{O}^{\Lambda,\mu \, \dag}_{D^{(\ast)}D^{\ast}}(\vec{P}) & = \sum_{\substack{\mu_1,\mu_2 \\ \vec{p_1},\vec{p_2}}} \mathcal{C}(\vec{P}\Lambda\mu;\vec{p_1}\Lambda_1\mu_1;\vec{p_2}\Lambda_2\mu_2) \\ 
	& \times \Omega_{D^{(\ast)}}^{\Lambda_1\mu_1 \, \dag}(\vec{p_1})\Omega_{D^{\ast}}^{\Lambda_2\mu_2 \, \dag}(\vec{p_2}) \, ,
\end{split}
\label{eq:memeop}
\end{equation}
with definite overall momentum $\vec{P}=\vec{p}_1+\vec{p}_2$, transforming in an irrep $\Lambda$ and row $\mu$ of the appropriate symmetry group. Here the single meson operators, $\Omega^{\Lambda_i\mu_i \, \dag}_{D^{(\ast)}}(\vec{p}_i)$, are variationally optimised fermion bilinears $\bar{\psi}_{\ell}\bm{\Gamma}\psi_c$ which interpolate the $D^{(\ast)}$ with momentum $\vec{p}_i$ in irrep $\Lambda_i$, and $\bm{\Gamma}$ is a product of $\gamma$-matrices and gauge covariant derivatives~\cite{Dudek:2010wm,Thomas:2011rh}. Up to three derivatives are used for mesons with $\vec{p}_i = 0$ and up to two derivatives for mesons with $\vec{p}_i \neq 0$, giving a large basis of operators in each irrep. The lattice Clebsch-Gordan coefficients, $\mathcal{C}$, couple the single meson operators to produce a $D^{(\ast)}D^{\ast}$ operator projected to $\Lambda = \Lambda_1 \otimes \Lambda_2$ with overall momentum $\vec{P}$, and all momenta related to $\vec{p}_1$ and $\vec{p}_2$ by allowed lattice rotations are summed over subject to the constraint that $\vec{P} = \vec{p}_1 + \vec{p}_2$. Isospin indices and the sum over isospin components to give overall isospin-0 have been suppressed in Eq.~\eqref{eq:memeop}. More details are given in Ref.~\cite{Dudek:2012gj} and pseudoscalar-vector scattering was first considered in this framework in Ref.~\cite{Woss:2018irj}.

At the light-quark masses chosen, the first three-hadron channel is $DD\pi$, given in Table~\ref{tab:masses}. This leads to a large window in energy where only $\DDst$ and $\DstDst$ are kinematically open. Included in the basis are all two hadron operators which in the absence of interactions correspond to a two hadron cm-frame energy, i.e. non-interacting energy, below the $DD\pi$ threshold. Also included are a limited amount of additional operators whose non interacting energies lie above the $DD\pi$ threshold to ensure a robust basis. In \cite{Cheung:2017tnt}, it was shown that the energy levels obtained using meson-meson operators were consistent within statistical uncertainty to the energy levels obtained using both meson-meson and local compact $cc\bar{q}\bar{q}$ operators in the doubly charmed sector with $I = 0, J^P = 1^+$ (see Figures 6 and 8 within Ref.~\cite{Cheung:2017tnt} ).\footnote{Ref. \cite{Cheung:2017tnt} did not relate the finite-volume energy levels to scattering amplitudes because only one relatively small volume was used and not enough energy levels were extracted to reliably determine the scattering amplitudes.} As such, we do not include local compact $cc\bar{q}\bar{q}$ operators in this study. The complete bases of operators used to interpolate the $D^{(\ast)}D^{\ast}$ states, both for zero and nonzero total momentum, can be found in Appendix \ref{sec:op_tables}.

Distillation vectors \cite{Peardon:2009gh} are utilized in order to obtain the necessary two-point functions to form the correlation matrix of Eq.~(\ref{eq:corr_matrix}). The distillation basis is formed using the lowest eigenmodes of the 3D gauge covariant Laplacian restricted to a time-slice $t$ of the lattice, which projects out the high frequency modes of the gauge fields. This projection creates smeared quark fields that are used in the interpolating operators, allowing reliable extraction of many states in the finite volume spectrum, such as those resembling mesons with various lattice-permitted momenta. Additionally, the use of distillation vectors allows for the required Wick contractions to be efficiently computed. The number of distillation vectors and the number of time sources are given in Table \ref{tab:latt_params}. The four irreps with overall nonzero momentum used three time sources. In order to reduce correlations, each non-zero momentum irrep used a different combination of three time sources from the four available.

\subsection{\label{sec:fvs}Finite Volume Spectra}

\begin{figure*}[!htbp]
\includegraphics[trim={0cm 0cm 5cm 0cm},clip,width=0.33\textwidth]{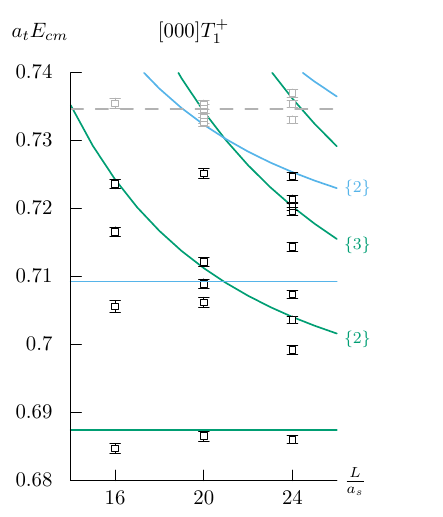}\hspace{-0.8cm}
\includegraphics[trim={0cm 0cm 5cm 0cm},clip,width=0.33\textwidth]{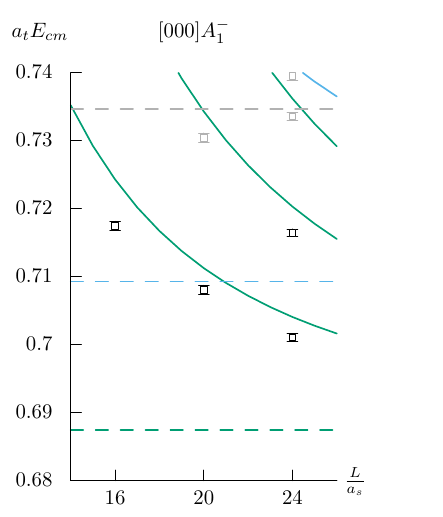}\hspace{-0.8cm}
\includegraphics[trim={0cm 0cm 5cm 0cm},clip,width=0.33\textwidth,scale=1.5]{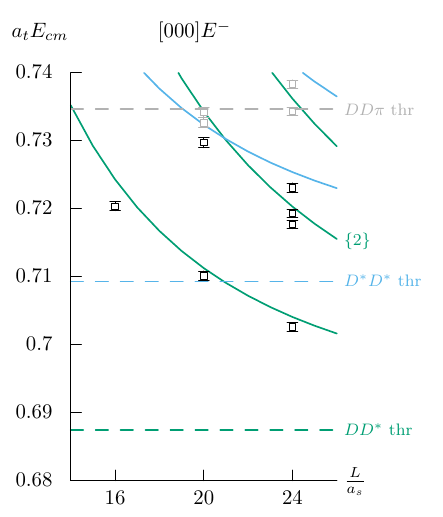}
\includegraphics[trim={0cm 0cm 5cm 0cm},clip,width=0.33\textwidth]{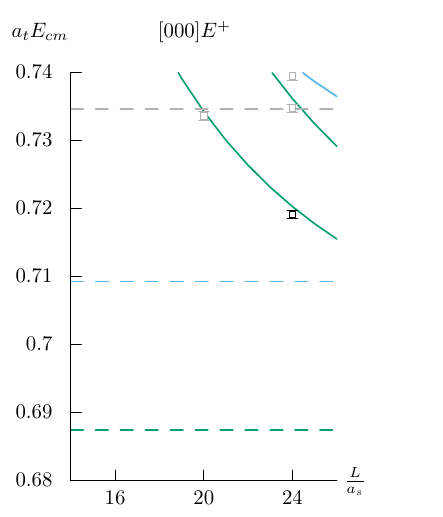}\hspace{-0.8cm}
\includegraphics[trim={0cm 0cm 5cm 0cm},clip,width=0.33\textwidth]{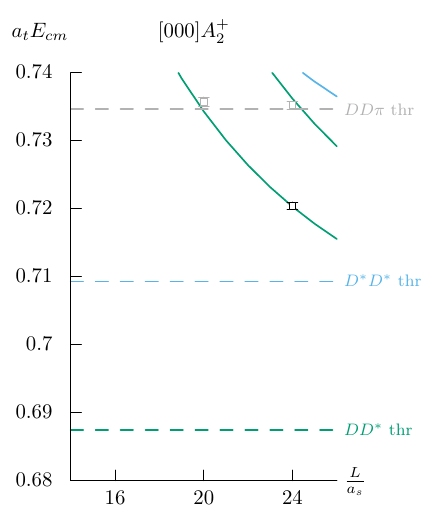}
	\caption{\label{fig:fve_rest_frames}The extracted finite-volume spectra for the $[000]\Lambda^P$ irreps. Points in black are used in the subsequent scattering analysis, while points in grey are not used. Solid curved lines denote the non interacting energies with the number of degenerate levels labelled in braces for $\DDst$ (green) and $\DstDst$ (blue) when the degeneracy is larger than one. Colored dashed lines denote relevant kinematic thresholds. The $D_0^*D$ threshold ($a_t E_{cm} = 0.73381$) has been ommitted due to its proximity to the $DD\pi$ threshold ($a_t E_{cm} = 0.73468$).}
\end{figure*}

\begin{figure*}[!htbp]
\includegraphics[trim={0cm 0cm 5cm 0cm},clip,width=0.33\textwidth]{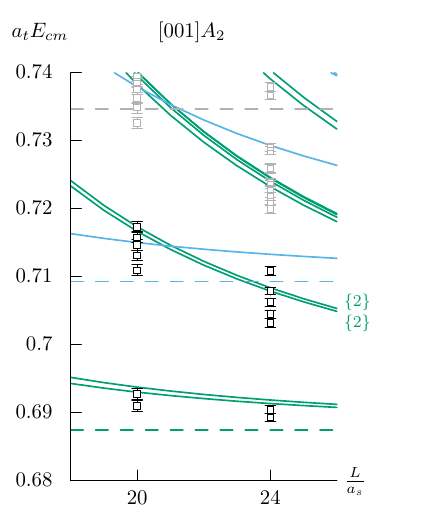}\hspace{-0.8cm}
\includegraphics[trim={0cm 0cm 5cm 0cm},clip,width=0.33\textwidth]{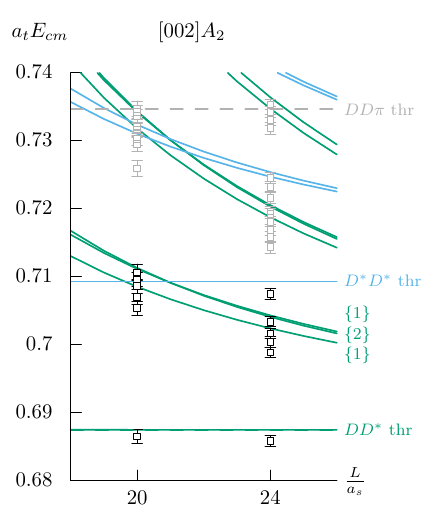}\hspace{5cm}
\includegraphics[trim={0cm 0cm 5cm 0cm},clip,width=0.33\textwidth]{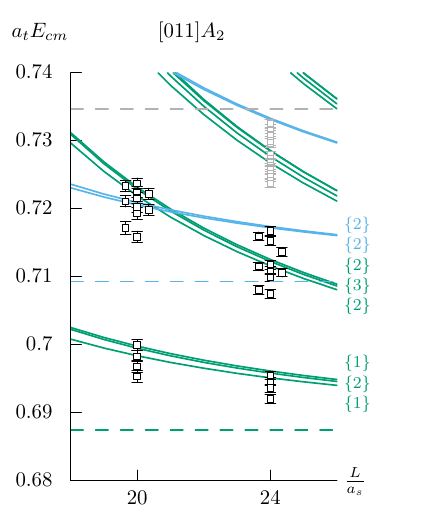}\hspace{-0.8cm}
\includegraphics[trim={0cm 0cm 5cm 0cm},clip,width=0.33\textwidth]{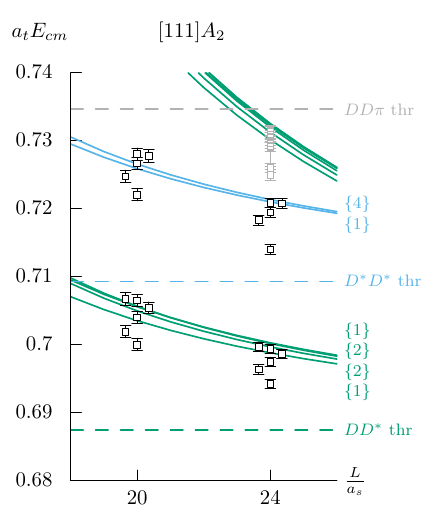}
\caption{\label{fig:fve_moving_frames}As Figure \ref{fig:fve_rest_frames}, but for the $[n_x n_y n_z]A_2$ irreps. Overlapping points have been displaced horizontally for clarity.}
\end{figure*}

Here the finite volume spectra in the $I=0$ doubly-charmed sector are presented. The irreps are labelled $[n_xn_yn_z]\Lambda^{(P)}$, where $[n_xn_yn_z] = \frac{2\pi}{L}(n_x,n_y,n_z) = \vec{P}$ is the quantized momentum, while $\Lambda$ labels the irrep with parity $P$. For $\vec{P} = 0$, we calculate the spectrum for the $T_1^+,A_1^-,E^-,E^+$ and $A_2^+$ irreps, using the basis of operators given in Table \ref{tab:rest_ops} of Appendix \ref{sec:op_tables}. Also calculated are the spectra for some irreps with $\vec{P} \neq 0$, namely $[001], [002], [011]$ and $[111] A_2$, using the basis of operators given in Table \ref{tab:inflight_ops} of Appendix \ref{sec:op_tables}. The use of the $A_2$ irreps allows us to constrain the $J^P=1^+$ scattering amplitudes in the full coupled-channel region, where the $[000]T^+_1$ irrep would offer only a small amount of constraint. The additional rest frame irreps $A_1^-,E^-,E^+$ and $A_2^+$ allow us to constrain other partial waves that subduce into $[000]T_1^+$ and the $A_2$ irreps. In Tables \ref{tab:pv_atrest}, \ref{tab:vv_atrest}, \ref{tab:pv_inflight} and \ref{tab:vv_inflight} in Appendix \ref{sec:sub_tables} we display the relevant partial wave subductions into these irreps for pseudoscalar-vector \cite{Woss:2018irj} and identical vector-vector meson scattering. 

Figures \ref{fig:fve_rest_frames} and \ref{fig:fve_moving_frames} show the extracted finite volume spectra for the irreps with $\vec{P} = 0$ and $\vec{P} \neq 0$, respectively. The relevant kinematic thresholds and non interacting energies, the expected cm-frame energies in the absence of any interaction, are labeled with their degeneracy. Points in black are those used in the subsequent scattering analysis, while those in grey are not used. In the $[000]T_1^+$ and $[002]A_2$ irreps, the ground state energies display significant shifts down from the $\DDst$ threshold, which is similar to previous lattice studies in $I=0$. In addition, the energy levels near the $\DstDst$ threshold have significant shifts downwards. The spectrum obtained on the $(L/a_s)^3 = 16^3$ lattice volume for $[000]T_1^+$ is consistent with the spectrum obtained in \cite{Cheung:2017tnt} both with and without local compact $cc\bar{q}\bar{q}$ operators. A similar picture is present in the $[n_xn_yn_z]A_2$ irreps. While the degeneracies of the non interacting levels are large in these irreps, we observe that some energy levels near $\DstDst$ threshold have significant shifts down away from the non interacting levels. We remark that all energy levels found in this calculation lie above the opening of the left hand cut due to one pion exchange between the $D$ and $D^{\ast}$, which begins at $a_t E_\cm = 0.68432$. Another left hand cut due to one pion exchange between $D^\ast$ and $D^\ast$ begins at $a_t E_\cm = 0.70591$. We will return to the discussion of the two left hand cuts in Section \ref{sec:systematics}.

\begin{figure*}[!htbp]
\includegraphics[trim={0cm 0.1cm 0cm 7.2cm},clip,width=0.95\textwidth,scale=0.5]{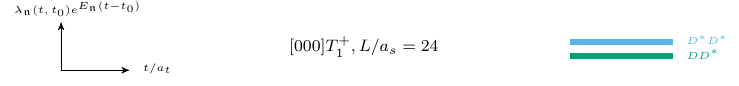}
\includegraphics[width=0.95\textwidth]{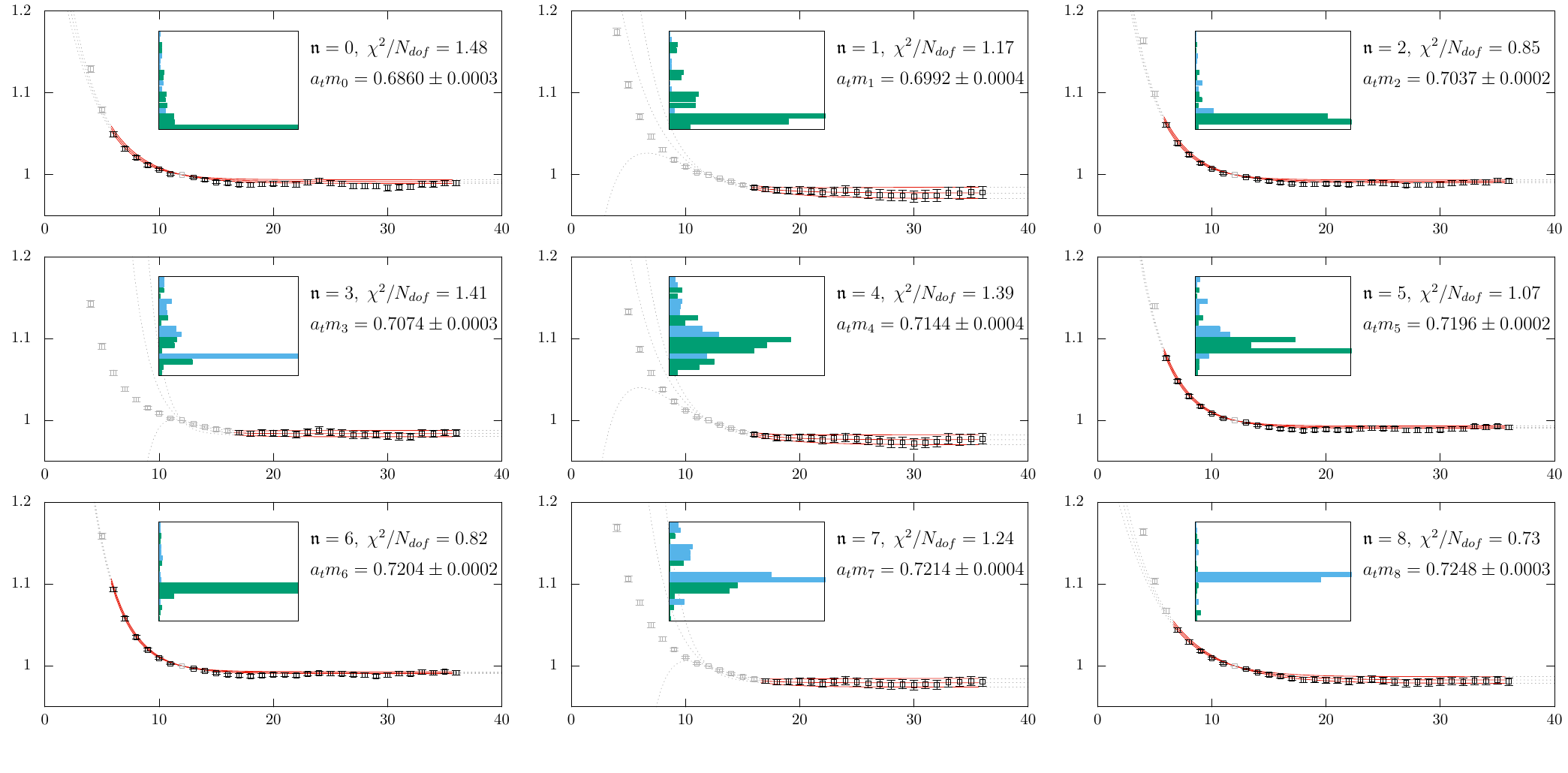}
	\caption{\label{fig:princ_corr} The principal correlators, plotted as $\lambda_{\mathfrak{n}}(t,t_0)e^{E_{\mathfrak{n}}(t-t_0)}$ with $t_0 = 12 a_t$, for the $[000]T_1^+$ irrep on the lattice of spatial extent $L / a_s = 24$. The result of the fit to the principle correlators is displayed in red, with the $\chi^2 / N_{\mathrm{dof}}$ and extracted $E_{\mathfrak{n}}$ from the fit. Points shown in gray were not used in the fit. Insets show histograms for the operator-state overlaps (normalized as explained in the text) for $\DDst$ (green) and $\DstDst$ (blue) operators ordered by increasing non interacting energy.}
\end{figure*}

In Figure \ref{fig:princ_corr}, we show the principal correlators with their corresponding fits, and the operator-state overlaps $Z_i^{\mathfrak{n}} =  \langle \mathfrak{n} \vert \mathcal{O}_i(0) \vert 0 \rangle$ for the first eight states in the $[000]T_1^+$ irrep, which are representative of the states extracted for the other irreps on each lattice volume. The normalization is fixed for each operator $i$ by requiring the largest observed magnitude of $Z_i^{\mathfrak{n}}$ over all $\mathfrak{n}$ to be 1.

As discussed in Ref. \cite{Wilson:2023anv}, there is a tension between the $D$ and $D^{\ast}$ meson masses on the three volumes due to some unaccounted for uncertainty (see Figure 25 of \cite{Wilson:2023anv}). We allow for this by adding an additional systematic uncertainty to the finite-volume energy levels using
\begin{equation}
	        a_t \delta E_\mathrm{latt} \rightarrow a_t \sqrt{\delta E_\mathrm{latt}^2 + \delta E^2_\mathrm{syst}},
\end{equation}
where $a_t\delta E_\mathrm{latt}$ is the uncertainty of the energy level and we set $a_t \delta E_\mathrm{syst} = 5 \times 10^{-4}$ as used in \cite{Wilson:2023anv}. The addition of this systematic error relieves the tension in the observed masses obtained from the relativistic dispersion fits of the $D$ and $D^{\ast}$ which may have been attributed to discretization effects or other additional sources of unknown systematic error. This added systematic error is propagated through to the uncertainty in the final scattering amplitudes. Further details can be found in Appendix A of \cite{Wilson:2023anv} in the context of charmonium scattering and \cite{Cheung:2020mql} in the context of $DK$ scattering. This tension has been observed in other lattice calculations using charm quarks \cite{Prelovsek:2020eiw,Piemonte:2019cbi,Xing:2022ijm} and various methods to relieve this tension can be employed. We view our approach as a relatively conservative method.

\section{\label{sec:scattering}Scattering Analysis}
In order to connect the finite volume energies with the infinite volume scattering amplitudes, we utilize the L\"{u}scher quantization condition~\cite{Luscher:1990ux}, extended to the coupled-channel scattering of hadrons of arbitrary mass and spin~\cite{Rummukainen:1995vs,Bedaque:2004kc,Kim:2005gf,He:2005ey,Lage:2009zv,Fu:2011xz,Leskovec:2012gb,Gockeler:2012yj,Hansen:2012tf,Briceno:2012yi,Guo:2012hv,Briceno:2014oea}
\begin{equation}
\mathrm{det}[{\bf 1}+ i{\pmb{\rho}} \cdot {\bf t} \cdot ({\bf 1} + i \pmb{ \cal M})] = 0
\label{eq:luscher}
\end{equation}
where $\pmb{\rho}(E_{cm})$ is a diagonal matrix of the phase space factors $2k/E_{cm}$ with $k$ the scattering momentum in the cm-frame, ${\bf t}(E_{cm})$ is the infinite volume scattering t-matrix, and $ \pmb{ {\cal M}}(E_{cm},L)$ is a known matrix dependent on the volume and irrep, and $E_{cm}$ is the energy in the cm-frame \cite{Briceno:2014oea}. The solutions of Eq.~(\ref{eq:luscher}), $\{E_{\mathfrak{n}}\}$, correspond to the finite volume energies for an appropriately parameterized ${\bf t}(E_{cm})$, which is related to the unitary scattering $S$-matrix via ${\bf S} = {\bf 1} + 2i\pmb{\rho}^{1/2}\cdot{\bf t}\cdot\pmb{\rho}^{1/2}$.

In coupled-channel scattering with multiple partial waves, the matrices in Eq.~(\ref{eq:luscher}) are in the space of all hadron-hadron scattering channels and partial waves. There are then multiple elements of the $t$-matrix that are unknown at any value of $E_{cm}$ and must be determined. A convenient parameterization for coupled-channel scattering is the $K$-matrix parameterization
\begin{widetext}
\begin{equation}
[t^{-1}(s)]_{\ell SJa,\ell 'S'Jb} = \frac{1}{(2k_a)^{\ell}}[K^{-1}(s)]_{\ell SJa,\ell 'S'Jb}\frac{1}{(2k_b)^{\ell '}} + \delta_{\ell \ell '}\delta_{SS'}I_{ab}(s)
\label{eq:tinv}
\end{equation}
\end{widetext}
where the indices run over orbital angular momentum $\ell$, spin $S$, angular momentum $J$ (the $t$-matrix is block diagonal in $J$), and hadron scattering channel $a,b$, with $s = E_{cm}^2$ and where the cm-frame momenta $k_{a,b}^\ell$ factors promote the expected behavior of partial wave amplitudes close to threshold. Unitarity of the $S$-matrix is ensured if the symmetric matrix $[K^{-1}(s)]_{\ell SJa,\ell' S'Jb}$ is real for real energies and the diagonal matrix $I_{ab}$ satisfies $\textrm{Im} I_{ab} = -\rho_a\delta_{ab}$ above threshold. The real part of $I_{ab}$ is either chosen to be zero (simple phase space), or a Chew-Mandelstam phase space is used, defined in terms of a dispersive integral as given in App. B of Ref.~\cite{Wilson:2014cna}. There is freedom in how to parameterise the elements of $[K^{-1}(s)]_{\ell SJa,\ell' S'Jb}$ and here we employ polynomials in $s = E_{cm}^2$,
\begin{equation}
K(s)_{\ell SJa,\ell'S'Jb} = \sum_{n}\gamma_{\ell SJa,\ell'S'Jb}^{(n)}s^n,
\label{eq:k_poly}
\end{equation}
where $\gamma_{\ell SJa,\ell'S'Jb}^{(n)}$ are real parameters to be determined. The parameters are determined using a correlated ${\chi}^2$ minimization of the difference between the finite volume energies computed on the lattice and the finite volume energies following from solutions of Eq.~(\ref{eq:luscher}) \cite{Wilson:2014cna,Woss:2020cmp}.

\subsection{\label{sec:elastic}$\DDst$ Scattering Below $\DstDst$ Threshold}
We first consider scattering where only the $\DDst$ channel is kinematically open, which we refer to as elastic scattering despite the inclusion of multiple partial waves. As is common when particles with non-zero spin are present, different combinations with the same $J^P$ can dynamically couple, such as \SLJ{3}{S}{1} and \SLJ{3}{D}{1} in $J^P=1^+$.
Ignoring the contribution of higher partial waves, Eq.~(\ref{eq:luscher}) reduces to a single equation for $S$-wave scattering and the $t$-matrix can be written in terms of a single real function, the scattering phase shift $\delta_0(E_{cm})$, via $t=\frac{1}{\rho}e^{i\delta_0}\sin(\delta_0)$. In principle, the S-wave phase shift could be extracted using the $[000]T_1^+$ irrep alone. However, many volumes would be required to have a set of energy levels that finely samples the energy dependence. We will therefore also consider irreps with overall non-zero momentum, but these have contributions from additional partial waves which must be constrained.

The possible partial wave configurations for $\DDst$ are given in Table \ref{tab:DDstPWs} up to $\ell = 3$ and the irreps to which they contribute are given in Tables \ref{tab:pv_atrest} and \ref{tab:pv_inflight} in Appendix \ref{sec:sub_tables}, which show partial wave contributions up to $\ell = 4$. As we wish to extract the $J^P = 1^+$ scattering amplitudes, we utilize the $[000]T_1^+$ irrep, where the \SLJ{3}{S}{1} and \SLJ{3}{D}{1,3} partial waves contribute. To increase the amount of constraint compared with using  $[000]T_1^+$ alone, we additionally utilize the $[001]$, $[002]$, $[011]$ and $[111]$ $A_2$ irreps, which also have contributions from \SLJ{3}{S}{1} and \SLJ{3}{D}{1}. However, these irreps with overall non zero momentum receive contributions from many other partial waves which must be constrained to extract the $J^P = 1^+$ scattering amplitude, notably the \SLJ{3}{P}{0,2}, \SLJ{3}{D}{2,3} and \SLJ{3}{F}{2,3,4} partial waves. To this end, we also utilize the $[000]A_1^-$ and $[000]E^-$ irreps, with lowest partial wave contribution \SLJ{3}{P}{0} and \SLJ{3}{P}{2}, respectively. 

In total, we use use 36 energy levels below $\DstDst$ threshold up to $a_tE_{\cm} = 0.705$.\footnote{The energy levels in $[000]A_2^+$ and $[000]E^+$, with lowest partial wave contribution \SLJ{3}{D}{3} and \SLJ{3}{D}{2}, respectively, lie above the energy region considered here, and as such are not utilized.} This is far enough below $\DstDst$ threshold to ensure that no energy levels which have dominant overlap with $\DstDst$ operators are included. We explicitly parameterize the partial waves \SLJ{3}{S}{1}, \SLJ{3}{P}{0,2} and \SLJ{3}{D}{1,2,3}. In this energy region, we assume that partial waves with $\ell \geq 3$ are negligible as they are expected to be highly suppressed due to threshold momentum factors in the absence of nearby singularities. \SLJ{3}{P}{1} is not considered, as this partial wave does not contribute to any of the irreps used. 

As mentioned previously, the \SLJ{3}{S}{1} and \SLJ{3}{D}{1} partial waves can dynamically mix in $J^P = 1^+$ which contributes an off-diagonal matrix element to Eq.~(\ref{eq:k_poly}). The scattering amplitude describing such a coupling could be expected to have a magnitude similar to P-wave based on the threshold momentum factors of Eq.~(\ref{eq:tinv}) and as such could be significant. However, we fix this and all other off-diagonal elements to zero for simplicity, which keeps Eq.~(\ref{eq:k_poly}) diagonal. This constraint will be relaxed in the full coupled-channel calculation of the subsequent section.

Utilizing a $K$-matrix parameterization of the form of Eq. (\ref{eq:k_poly}), we find that a reasonable description can be obtained with a constant and linear term in $s$ for \SLJ{3}{S}{1}, and constant elements for others,

\begin{widetext}
\center
\renewcommand{\arraystretch}{1.5}
\begin{small}
\begin{tabular}{rll}
$\gamma^{(0)}_{\SLJ{3}{S}{1}} = $ & $(8.89 \pm 0.81 \pm 1.16) $ & \multirow{7}{*}{ $\begin{bmatrix*}[r]   1.00 &  -0.97 &   0.60 &   0.55 &   0.35 &   0.25 &   0.13\\
&  1.00 &  -0.54 &  -0.49 &  -0.33 &  -0.24 &  -0.09\\
&&  1.00 &   0.74 &   0.44 &   0.17 &   0.29\\
&&&  1.00 &   0.49 &   0.22 &   0.17\\
&&&&  1.00 &   0.28 &  -0.62\\
&&&&&  1.00 &  -0.30\\
&&&&&&  1.00\end{bmatrix*}$ } \\
$\gamma^{(1)}_{\SLJ{3}{S}{1}} = $ & $(-241 \pm 35 \pm 70) \cdot a_t^2$ & \\
$\gamma^{(0)}_{\SLJ{3}{P}{0}} = $ & $(181 \pm 15 \pm 24) \cdot a_t^2$ & \\
$\gamma^{(0)}_{\SLJ{3}{P}{2}} = $ & $(76 \pm 9 \pm 17) \cdot a_t^2$ & \\
$\gamma^{(0)}_{\SLJ{3}{D}{1}} = $ & $(1218 \pm 390 \pm 719) \cdot a_t^4$ & \\
$\gamma^{(0)}_{\SLJ{3}{D}{2}} = $ & $(-1408 \pm 761 \pm 288) \cdot a_t^4$ & \\
$\gamma^{(0)}_{\SLJ{3}{D}{3}} = $ & $(-410 \pm 248 \pm 414) \cdot a_t^4$ & \\[1.3ex]
&\multicolumn{2}{l}{ $\chi^2/ N_\mathrm{dof} = \frac{39.7}{36-7} = 1.37$\,,}
\end{tabular}
\end{small}
\begin{equation}
\label{eq:elastic_params}
\end{equation}
\end{widetext}
where we have suppressed the index on the hadronic scattering channel since only $\DDst$ scattering is present. For brevity, we have labeled the matrix elements $\gamma_{\ell SJa,\ell'S'Jb}^{(n)}$ in Eq.~(\ref{eq:k_poly}) as $\gamma_{\SLJ{2S+1}{\ell}{J}}^{(n)}$. In this parameterization and all others that follow, any parameter not listed is implicitly fixed to zero. The first errors are statistical while the second are systematic errors arising from the envelope of parameter values obtained by performing separate $\chi^2$-minimizations using the 1$\sigma$ deviations from the central values of the $D$ and $D^{\ast}$ mesons masses and anisotropy. In the square brackets we display the correlation between the individual parameters.

\begin{figure}
\includegraphics[width=0.95\columnwidth]{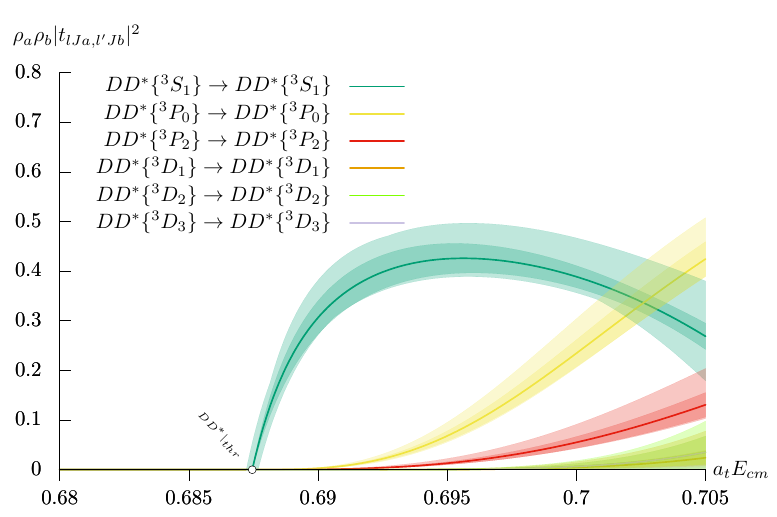}
\includegraphics[trim={0cm 0cm 0cm 6.5cm},clip,width=0.95\columnwidth]{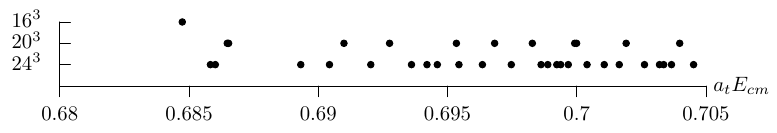}
	\caption{\label{fig:amps_elastic}(Top) $\DDst$  elastic scattering amplitudes in the \SLJ{3}{S}{1}, \SLJ{3}{P}{0,2} and \SLJ{3}{D}{1,2,3} partial waves for the parameterization given by Eq.~(\ref{eq:elastic_params}). Inner shaded bands reflect the statistical uncertainty on the parameterization and outer bands represent the systematic uncertainty obtained by varying the mass of the $D$ and $D^{\ast}$ mesons and the anisotropy as described in the text. (Bottom) The energy levels used to constrain the scattering amplitudes.}
\end{figure}
Fig. \ref{fig:amps_elastic} shows the scattering amplitudes from the $K$-matrix parameterization of Eq.~(\ref{eq:elastic_params}). We observe a rapid turn on of the $S$-wave amplitude at $\DDst$ threshold, suggestive of a pole in the $t$-matrix close to threshold. The $P$-wave amplitudes are significantly non-zero. However, we observe that \SLJ{3}{P}{0} and \SLJ{3}{P}{2} differ which can perhaps be anticipated from the larger downward shifts for the lowest levels in $[000]A_1^-$ compared with $[000]E^-$. This will be further discussed in Section \ref{sec:coupled}. The $D$-wave amplitudes are found to be negligible.

The $S$-wave $t$-matrix (in the absence of higher partial waves) can be recast to the form $t = \frac{E_{\cm}}{2}\frac{1}{k\cot\delta_0-ik}$, which makes it obvious that a pole singularity will manifest through an intersection of $k\cot\delta_0$ with $\pm ik$ at an energy below threshold. An intersection of $k\cot\delta_0$ with $-ik$ corresponds to a pole on the real axis below threshold on the physical sheet, i.e. a bound state. In contrast, an intersection of $k\cot\delta_0$ with $ik$ corresponds to a pole below threshold on the unphysical sheet, i.e. a virtual bound state.

\begin{figure}
\includegraphics[width=0.98\columnwidth]{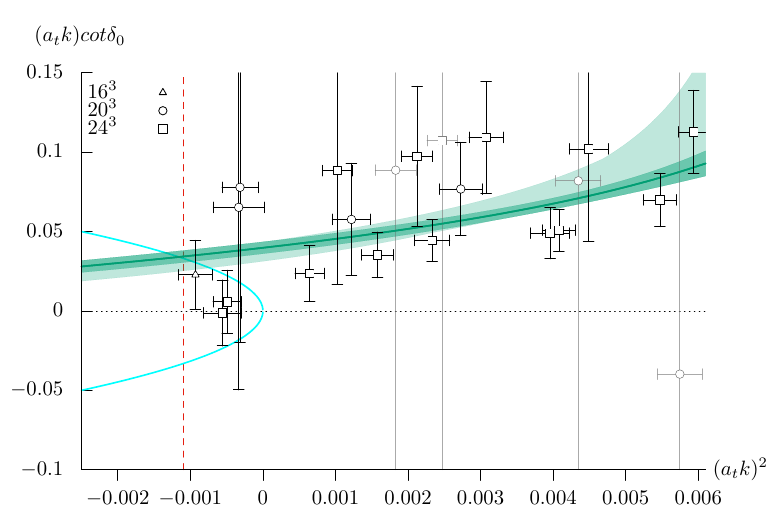}
\includegraphics[width=0.95\columnwidth]{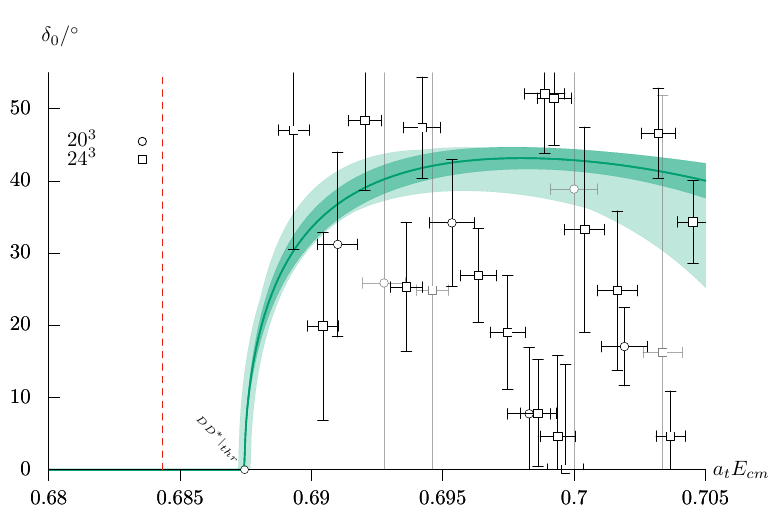}
	\caption{\label{fig:kcotdelta}(Top) The $S$-wave amplitude from Eq.~(\ref{eq:elastic_params}), shown as $(a_t k)\cot\delta_0$ (green), along with $\pm a_t|k|$ (blue) for $k^2<0$. The inner shaded band reflects the statistical uncertainty while the outer band represents the systematic uncertainty obtained by varying the mass of the $D$ and $D^{\ast}$ mesons and the anisotropy as described in the text. (Bottom) The corresponding phase shift, $\delta_0$. Discrete points are determined from individual energy levels using the determinant condition Eq.~(\ref{eq:luscher}). Those with large statistical error due to their proximity to non interacting levels have been plotted in grey to lessen their visual impact. The vertical dashed red line in both plots denotes the position of the beginning of the left hand cut due to one pion exchange between the $D$ and $D^\ast$.}
\end{figure}

Fig.~\ref{fig:kcotdelta} (top) shows a plot of $k\cot\delta_0$ as a function of $k^2$ for the extracted $S$-wave phase shift (bottom). We observe a clear intersection of the $k\cot\delta_0$ curve with $ik$, corresponding to a virtual bound state. The position of the virtual bound state using this elastic analysis will be discussed in Section \ref{subsec:elastic_poles}. In the next section we will expand upon the determination of the scattering amplitudes and the virtual bound state using a coupled-channel analysis.

\subsection{\label{sec:coupled}Coupled-channel $\DDst$, $\DstDst$ scattering}
Building upon the parameterization of elastic $\DDst$ scattering, we now consider higher energies where both the $\DDst$ and $\DstDst$ channels are kinematically accessible and so we consider coupled-channel $\DDst$-$\DstDst$ scattering. In this higher-energy region, partial waves that are suppressed close to threshold have more phase space and can become important.
In order to have sufficient constraint, 109 energy levels are used from both the rest-frame irreps and those with non-zero overall momentum, up to energies of $a_tE_{cm} = 0.73$. This is far enough below the $DD\pi$ threshold to ensure that no three-body effects arise\footnote{This also ensures that no effects from $D_0^{\ast}D$ scattering contribute, as the $D_0^{\ast}D$ threshold opens just below the $DD\pi$ threshold.}.

On each volume large correlations are observed between the energy levels. This has been seen before on these lattices in calculations with charm quarks, as described in Refs.~\cite{Cheung:2020mql,Wilson:2023anv}. It is possible that the correlations between energy levels are not reliably determined given the number of gauge configurations used ($\approx 500$ in the case of the $24^3$ lattice) and the number of independent numbers in the correlation matrix (which grows like $N_{\mathrm{levels}}^2$). As found previously, some method for taming these large correlations is necessary in the coupled-channel case, which is not required in the elastic analysis as the number of required energy levels to provide constraint for the scattering amplitudes is significantly smaller there.

In this study, we use ``singular value limits'' in the $\chi^2$ minimization. Here the replacement $\lambda_i \rightarrow \mathrm{max}(\lambda_i,\sigma)$ for the eigenvalues, $\lambda_i$, of the correlation matrix is made, where $\sigma = \tau \lambda_1$, $\tau$ is the singular value limit and $\lambda_1$ is the largest eigenvalue of the correlation matrix. This has the effect of lowering the correlations between finite volume energies, but increases the uncertainties~\cite{Dowdall:2019bea}. An alternative is to use ``singular value cuts'', wherein modes corresponding to eigenvalues of the correlation matrix below $\sigma$ are removed, this was used in Refs.~\cite{Cheung:2020mql,Wilson:2023anv}. In practical terms, the approaches are similar, however for a given value of the cutoff $\tau$, ``limiting'' is found to be less conservative than ``cutting''. We choose a value of $\tau = 10^{-2}$. 
For further discussion of the use of singular value limits, see Appendix~\ref{sec:svdlims}. 

In this coupled-channel analysis, we seek to determine the $J^P = 1^+$ scattering amplitudes of both $\DDst$ and $\DstDst$. These amplitudes can be simultaneously extracted by using energy levels around and above $\DstDst$ threshold in the same irreps utilized in the elastic analysis, as energy levels dominated by $\DstDst$ operators begin to appear just below $\DstDst$ threshold (see Figure \ref{fig:princ_corr}). The high degeneracy of non-interacting levels in the upper part of the spectra for the irreps utilized mirrors the large number of partial waves that are present, and these must be accounted for when extracting the $J^P = 1^+$ $\DDst$ and $\DstDst$ scattering amplitudes. 

In the elastic analysis, we considered all partial waves for $\DDst$ with $\ell \leq 2$ that contribute to the irreps utilized. These partial waves are also considered in this coupled-channel analysis. As we now seek to constrain the scattering amplitudes up to $a_t E_{cm} = 0.73$, additional energy levels in the $[000]A_1^-$ and $[000]E^-$ irreps are available, which we can use to further constrain the $\DDst \SLJc{3}{P}{0,2}$ amplitudes. Additionally, the energy levels in the $[000]A_2^+$ and $[000]E^+$ irreps can be used to further constrain the \SLJ{3}{D}{2,3} partial waves. We also choose to include \SLJ{3}{F}{2}, which contributes to the $[000]E^-$ and all $A_2$ irreps with non zero momentum, to account for the large degeneracy of non-interacting $\DDst$ levels in the upper part of the spectrum.

We must also determine $\DstDst$ amplitudes in these irreps. There are fewer $\DstDst$ combinations due to Bose symmetry as discussed in Section~\ref{sec:symm}; see Table~\ref{tab:DstDstPWs} and the irreps to which they contribute in Tables~\ref{tab:vv_atrest} and \ref{tab:vv_inflight}. In contrast to the $\DDst$ partial waves, we do not include all $\DstDst$ partial waves with $\ell \leq 2$. Of the allowed $\DstDst$ partial waves, we consider \SLJ{3}{S}{1}, \SLJ{3}{D}{1}, \SLJ{5}{P}{2} and \SLJ{5}{F}{2}. Similarly to $\DDst$, the $\DstDst$ \SLJ{3}{S}{1} and \SLJ{3}{D}{1} partial waves contribute to $[000]T_1^+$ and all non-zero momentum $A_2$ irreps, while the \SLJ{5}{P}{2} and \SLJ{5}{F}{2} partial waves contribute to $[000]E^-$ and all non-zero momentum $A_2$ irreps. The choice to include \SLJ{5}{P}{2} is motivated by the magnitude of the scattering amplitude of $\DDst$ in \SLJ{3}{P}{2}, which was observed to be significant in the elastic analysis of Section~\ref{sec:elastic} (see Figure~\ref{fig:amps_elastic}). A similar effect could be expected of the $\DstDst$ \SLJ{5}{P}{2} partial wave. Additionally, a fourth amplitude is required to satisfy the quadruple degeneracy in the upper part of the spectrum in the $[111]A_2$ irrep (see Figure \ref{fig:fve_moving_frames}). In Appendix \ref{sec:cc_hpw}, we show there is a freedom in the choice of which partial wave to parameterize for, which we choose to be the partial wave \SLJ{5}{F}{2}.

Not accounted for are the $\DstDst$ partial waves \SLJ{3}{D}{2,3}, \SLJ{5}{P}{3}, and \SLJ{5}{F}{3,4,5}. In Appendix \ref{sec:cc_hpw}, we justify neglecting the partial waves, and show they do not have an effect on the $J^P = 1^+$ scattering amplitudes. The $\DstDst$ \SLJ{1}{P}{1}, \SLJ{5}{P}{1} and \SLJ{5}{F}{1} partial waves are not considered as these do not contribute to the irreps utilized.

Analogously to the elastic case, we utilize a $K$-matrix parameterization of the form of Eq.~(\ref{eq:k_poly}) to determine simultaneously the $J^P=1^+$ amplitude and the smaller partial waves in $J^P = 0^-,2^-,2^+$ and $3^+$. Performing this amplitude determination, we find a good description of the spectra using a constant plus linear term in $s$ for the $\DDst$ and $\DstDst$ $S$-wave amplitudes and constants for all others,\footnote{Correlations are present between the parameters in all $J^P$. However, for brevity we have chosen to show only the larger diagonal blocks. Parameter correlations between different $J^P$ are typically much smaller than those within a given $J^P$.}
\begin{widetext}
\begin{small}
\begin{center}
\renewcommand{\arraystretch}{1.4}
\begin{tabular}{rll}
$\gamma^{(0)}_{\DDst \SLJc{3}{S}{1} \rightarrow \DDst \SLJc{3}{S}{1}} = $ & $(6.68 \pm 0.53 \pm 0.31)$ & \multirow{7}{*}{ $\begin{bmatrix*}[r]   1.00 &  -0.80 &   0.33 &  -0.12 &   0.25 &   0.07 &   0.10\\
&  1.00 &   0.19 &   0.23 &  -0.27 &  -0.00 &  -0.04\\
&&  1.00 &   0.27 &  -0.04 &   0.06 &   0.09\\
&&&  1.00 &  -0.95 &   0.04 &   0.06\\
&&&&  1.00 &  -0.03 &  -0.02\\
&&&&&  1.00 &   0.03\\
&&&&&&  1.00\end{bmatrix*}$ } \\ 
$\gamma^{(1)}_{\DDst \SLJc{3}{S}{1} \rightarrow \DDst \SLJc{3}{S}{1}} = $ & $(-59 \pm 13 \pm 21) \cdot a_t^2$ & \\
$\gamma^{(0)}_{\DDst \SLJc{3}{S}{1} \rightarrow \DstDst \SLJc{3}{S}{1}} = $ & $(4.11 \pm 0.51 \pm 0.94)$ & \\
$\gamma^{(0)}_{\DstDst \SLJc{3}{S}{1} \rightarrow \DstDst \SLJc{3}{S}{1}} = $ & $(12.7 \pm 2.5 \pm 2.8)$ & \\
$\gamma^{(1)}_{\DstDst \SLJc{3}{S}{1} \rightarrow \DstDst \SLJc{3}{S}{1}} = $ & $(-122 \pm 53 \pm 71) \cdot a_t^2$ & \\
$\gamma^{(0)}_{\DDst \SLJc{3}{D}{1} \rightarrow \DDst \SLJc{3}{D}{1}}= $ & $(149 \pm 63 \pm 42) \cdot a_t^4$ & \\
$\gamma^{(0)}_{\DstDst \SLJc{3}{D}{1} \rightarrow \DstDst \SLJc{3}{D}{1}} = $ & $(-504 \pm 178 \pm 1095) \cdot a_t^4$ & \\
\end{tabular}
\end{center}
\begin{center}
\renewcommand{\arraystretch}{1.4}
\begin{tabular}{rll}
$\gamma^{(0)}_{\DDst \SLJc{3}{P}{0} \rightarrow \DDst \SLJc{3}{P}{0}} = $ & $(66 \pm 5 \pm 14) \cdot a_t^2$ & \multirow{5}{*}{ $\begin{bmatrix*}[r]   1.00 &  -0.01 &   0.14 &   0.30 &   0.08\\
&  1.00 &   0.14 &  -0.73 &   0.05\\
&&  1.00 &  -0.02 &  -0.18\\
&&&  1.00 &  -0.00\\
&&&&  1.00\end{bmatrix*}$ } \\ 
$\gamma^{(0)}_{\DDst \SLJc{3}{P}{2} \rightarrow \DDst \SLJc{3}{P}{2}} = $ & $(29 \pm 3.9 \pm 3.9) \cdot a_t^2$ & \\
$\gamma^{(0)}_{\DstDst \SLJc{5}{P}{2} \rightarrow \DstDst \SLJc{5}{P}{2}} = $ & $(83 \pm 11 \pm 4) \cdot a_t^2$ & \\
$\gamma^{(0)}_{\DDst \SLJc{3}{F}{2} \rightarrow \DDst \SLJc{3}{F}{2}} = $ & $(-126 \pm 1095 \pm 3181) \cdot a_t^6$ & \\
$\gamma^{(0)}_{\DstDst \SLJc{5}{F}{2} \rightarrow \DstDst \SLJc{5}{F}{2}} = $ & $(-25297 \pm 15400 \pm 15133) \cdot a_t^6$ & \\
\end{tabular}
\end{center}
\begin{center}
\renewcommand{\arraystretch}{1.4}
\begin{tabular}{rll}
$\gamma^{(0)}_{\DDst \SLJc{3}{D}{2} \rightarrow \DDst \SLJc{3}{D}{2}} = $ & $(82 \pm 64 \pm 93) \cdot a_t^4$ & \multirow{2}{*}{ $\begin{bmatrix*}[r]   1.00 &   0.07\\
&  1.00\end{bmatrix*}$ } \\ 
$\gamma^{(0)}_{\DDst \SLJc{3}{D}{3} \rightarrow \DDst \SLJc{3}{D}{3}} = $ & $(-76 \pm 38 \pm 83) \cdot a_t^4$ & \\[1.3ex]
&\multicolumn{2}{l}{ $\chi^2/ N_\mathrm{dof} = \frac{121.3}{109-14} = 1.28$\,,}
\end{tabular}
\end{center}
\end{small}
\begin{equation}
\label{eq:ref_param}
\end{equation}
\end{widetext}
where the first and second uncertainties are as described in Section \ref{sec:elastic}. We refer to Eq.~(\ref{eq:ref_param}) as the \textit{reference} parameterization. Very little deviation is observed in the parameters describing the partial waves in $J^P = 0^-,2^-,2^+$ and $3^+$ from this determination to when they are determined independently from rest-frame irreps alone.  Of particular importance in the reference parameterization is the nonzero value of the matrix element $\gamma^{(0)}_{\DDst \SLJc{3}{S}{1} \rightarrow \DstDst \SLJc{3}{S}{1}}$. This matrix element allows the $\DDst\SLJc{3}{S}{1}$ and $\DstDst\SLJc{3}{S}{1}$ channels to mix and is critical to obtaining a parameterization suitable for replicating the finite volume spectra, which will be demonstrated in Section \ref{subsec:coupling}.

\begin{figure*}[!htbp]
	\includegraphics[trim={0cm 0cm 5cm 0cm},clip,width=0.33\textwidth]{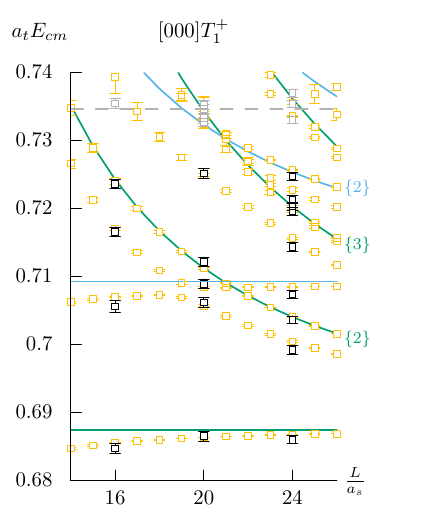}\hspace{-0.8cm}
	\includegraphics[trim={0cm 0cm 5cm 0cm},clip,width=0.33\textwidth]{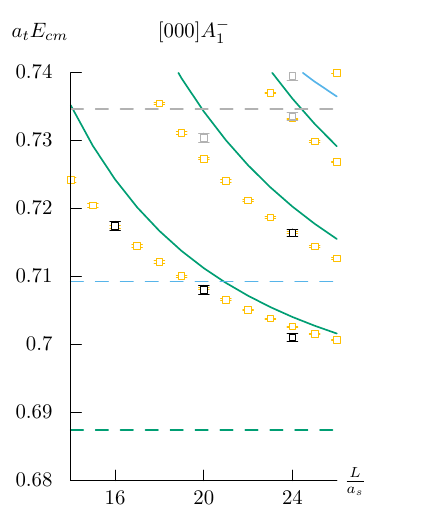}\hspace{-0.8cm}
	\includegraphics[trim={0cm 0cm 5cm 0cm},clip,width=0.33\textwidth,scale=1.5]{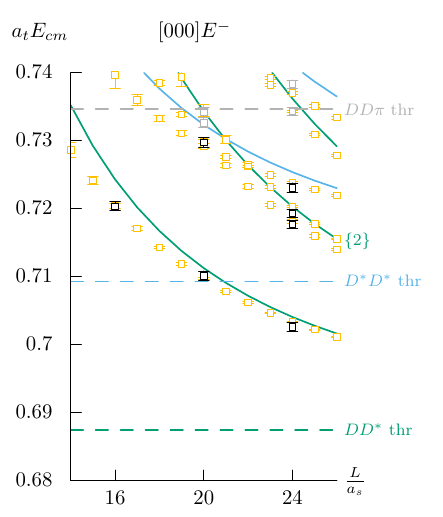}\hspace{-0.8cm}
	\includegraphics[trim={0cm 0cm 5cm 0cm},clip,width=0.33\textwidth]{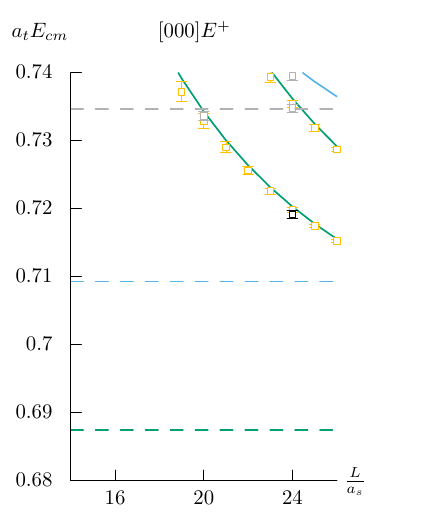}\hspace{-0.8cm}
	\includegraphics[trim={0cm 0cm 5cm 0cm},clip,width=0.33\textwidth]{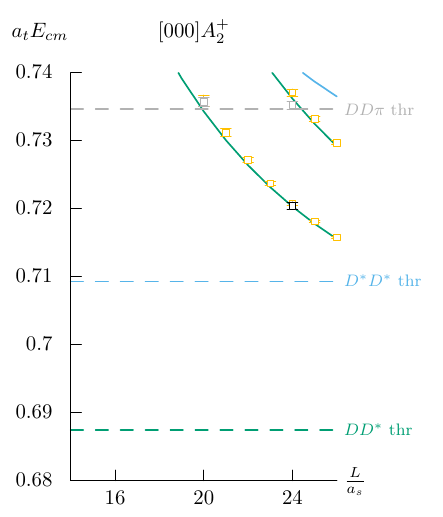}
	\caption{\label{fig:fvs_rest_frames}As Fig. \ref{fig:fve_rest_frames}, but with the finite volume spectra obtained from Eq.~(\ref{eq:luscher}) using the reference parameterization of Eq.~(\ref{eq:ref_param}) given by yellow points. }
\end{figure*}

\begin{figure*}[!htbp]
	\includegraphics[trim={0cm 0cm 5cm 0cm},clip,width=0.33\textwidth]{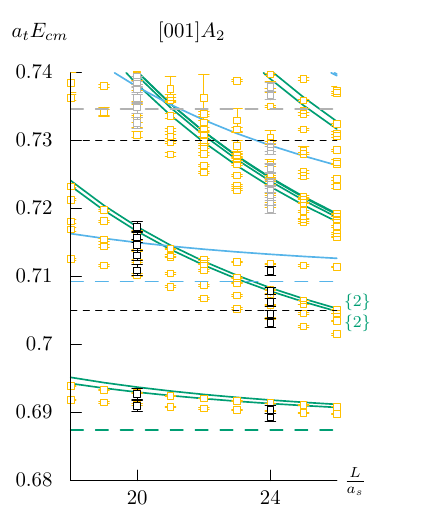}\hspace{-0.8cm}
	\includegraphics[trim={0cm 0cm 5cm 0cm},clip,width=0.33\textwidth]{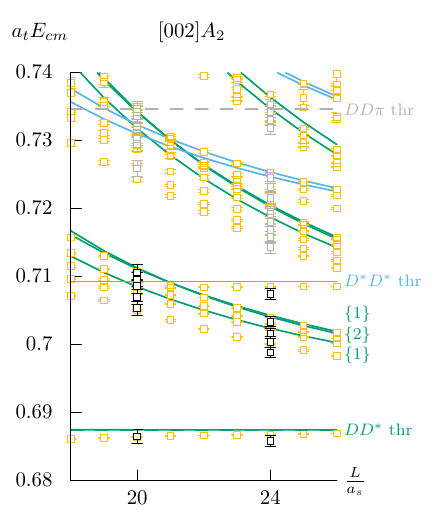}\hspace{5cm}
	\includegraphics[trim={0cm 0cm 5cm 0cm},clip,width=0.33\textwidth]{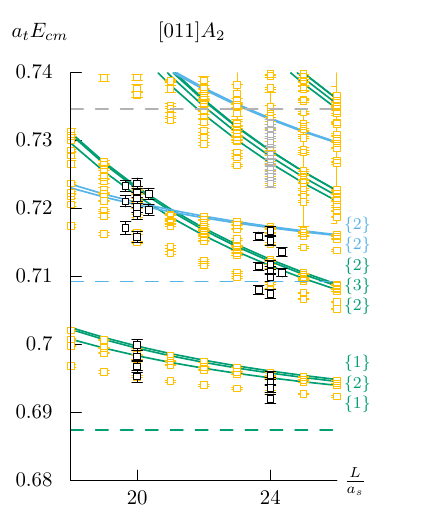}\hspace{-0.8cm}
	\includegraphics[trim={0cm 0cm 5cm 0cm},clip,width=0.33\textwidth]{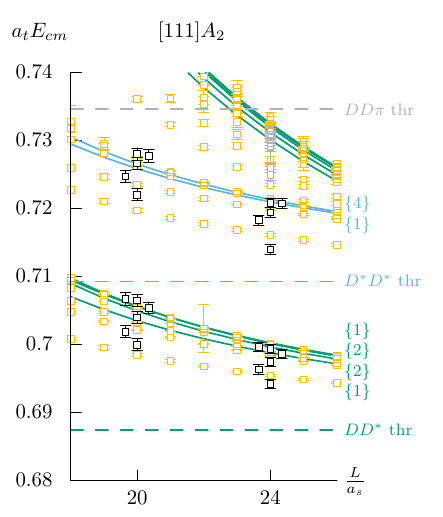}
	\caption{\label{fig:fvs_moving_frames}As Figure \ref{fig:fvs_rest_frames}, but for the $[n_x n_y n_z]A_2$ irreps. Overlapping points have been displaced horizontally for clarity.}
\end{figure*}

Figures \ref{fig:fvs_rest_frames} and \ref{fig:fvs_moving_frames} display the finite volume spectra obtained from Eq.~(\ref{eq:luscher}) using the reference parameterization of Eq.~(\ref{eq:ref_param}) for the at rest and in flight irreps, respectively. The lattice energies are well replicated across all irreps. An avoided level crossing occurs in the $[000]T_1^+$ irrep just below $\DstDst$ threshold, indicating some coupling between the channels.

In order to reduce bias caused by the use of a single parameterization, we also consider 13 additional parameterizations listed in Table \ref{tab:param_variations}, some of which include the effects of $S$ and $D$-wave mixing for both $\DDst$ and $\DstDst$. Out of the 14 parameterization variations, two were found to give $\chi^2 / N_{\mathrm{dof}} > 1.5$ and so were not used in the subsequent analysis.

\begin{table*}[!htbp]
	\caption{\label{tab:param_variations}Variations of the $K$-matrix parameterization used in coupled-channel scattering. The entries of the table display the order of the polynomial in $s$ utilized. $-$ denotes that the parameter was fixed to 0 for that parameterization. The reference parameterization Eq.~(\ref{eq:ref_param}) is denoted in bold with $\ast$. $\dag$ denotes a parameterization that is \emph{not} used in the subsequent analysis. All other waves were parametrised in the same way as in the reference amplitude, Eq.~(\ref{eq:ref_param}), using a polynomial of order 0 and a Chew-Mandelstam (Chew-Man.) phase space, so they are not included in the table.}
\begin{ruledtabular}
\begin{tabular}{c|c|c|c|c|c|c|c|c}
	\multirow{2}{4em}{Phase Space} & $DD^{\ast}$\SLJc{3}{S}{1} & $DD^{\ast}$\SLJc{3}{S}{1} & $D^{\ast}D^{\ast}$\SLJc{3}{S}{1} & $DD^{\ast}$\SLJc{3}{S}{1} & $D^{\ast}D^{\ast}$\SLJc{3}{S}{1} & $DD^{\ast}$\SLJc{3}{D}{1} & $D^{\ast}D^{\ast}$\SLJc{3}{D}{1} & \multirow{2}{4em}{$\chi^2 / N_{\mathrm{dof}}$} \\
	& $\rightarrow$$DD^{\ast}$\SLJc{3}{S}{1} & $\rightarrow$$D^{\ast}D^{\ast}$\SLJc{3}{S}{1} & $\rightarrow$$D^{\ast}D^{\ast}$\SLJc{3}{S}{1} & $\rightarrow$$DD^{\ast}$\SLJc{3}{D}{1} & $\rightarrow$$D^{\ast}D^{\ast}$\SLJc{3}{D}{1} & $\rightarrow$$DD^{\ast}$\SLJc{3}{D}{1} & $\rightarrow$$D^{\ast}D^{\ast}$\SLJc{3}{D}{1} \\
\hline
	\multirow{7}{4em}{Chew-Man.} & 0 & 0 & 1 & $-$ & $-$ & 0 & 0 & 1.45 \\
	& \textbf{1} &  \textbf{0} &  \textbf{1} &  $\bm{-}$ & $\bm{-}$ &  \textbf{0} &  \textbf{0} & $\bm{1.28^*}$ \\
	& 0 & 1 & 1 & $-$ & $-$ & 0 & 0 & 1.36 \\
	& 1 & 1 & 0 & $-$ & $-$ & 0 & 0 & 1.30 \\
	& 1 & 1 & 1 & $-$ & $-$ & 0 & 0 & 1.29\\
	& 1 & 0 & 1 & 0 & 0 & 0 & 0 & 1.23 \\
	& 1 & 0 & 1 & 0 & 0 & $-$ & $-$ & 1.31 \\
\hline
	\multirow{5}{4em}{Simple} & 0 & 0 & 1 & $-$ & $-$ & 0 & 0 & $1.63^{\dag}$ \\
	& 1 & 0 & 1 & $-$ & $-$ & 0 & 0 & 1.30 \\
	& 0 & 1 & 1 & $-$ & $-$ & 0 & 0 & $1.53^{\dag}$ \\
	& 1 & 1 & 0 & $-$ & $-$ & 0 & 0 & 1.33  \\
	& 1 & 1 & 1 & $-$ & $-$ & 0 & 0 & 1.31 \\
	& 1 & 0 & 1 & 0 & 0 & 0 & 0 & 1.25 \\
	& 1 & 0 & 1 & 0 & 0 & $-$ & $-$ & 1.32 \\
\end{tabular}
\end{ruledtabular}
\end{table*}

\begin{figure*}[!htbp]
\includegraphics[width=0.95\columnwidth]{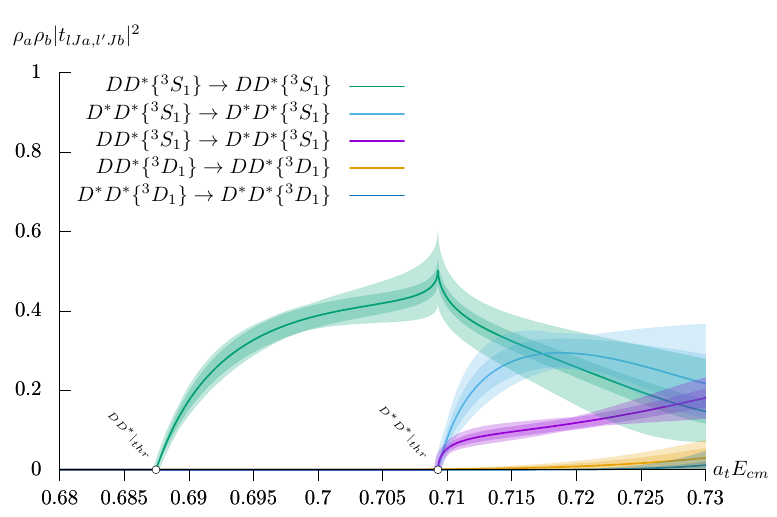}
\includegraphics[width=0.95\columnwidth]{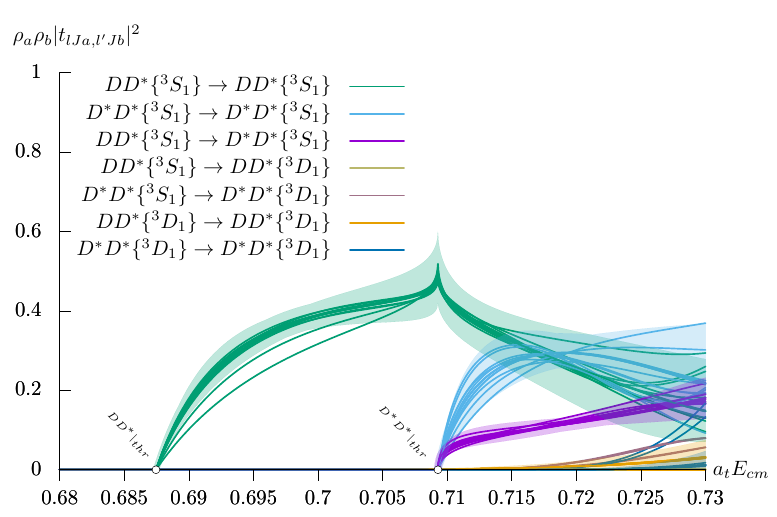}

\includegraphics[width=0.95\columnwidth]{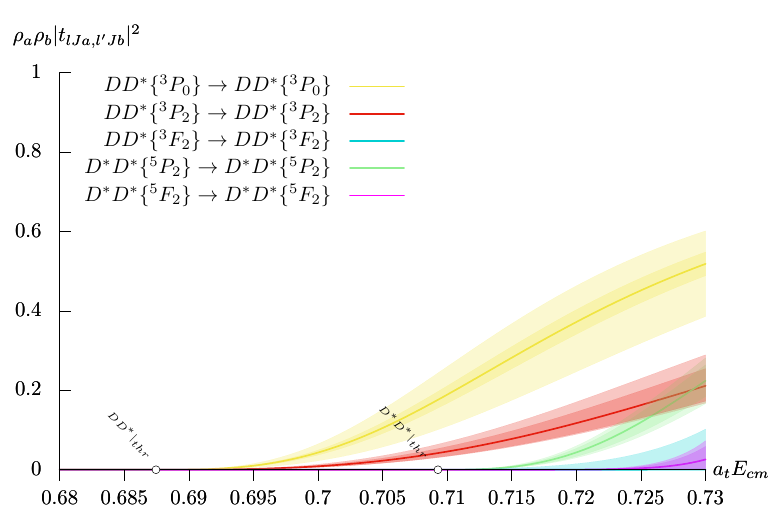}
\includegraphics[width=0.95\columnwidth]{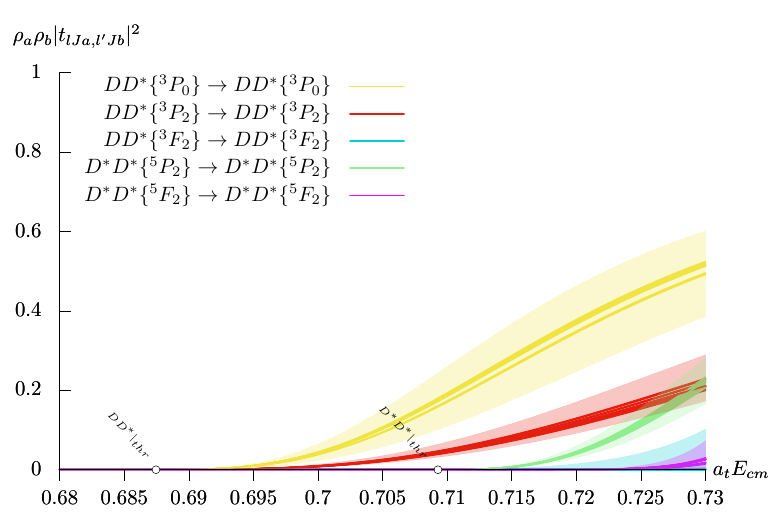}

\includegraphics[width=0.95\columnwidth]{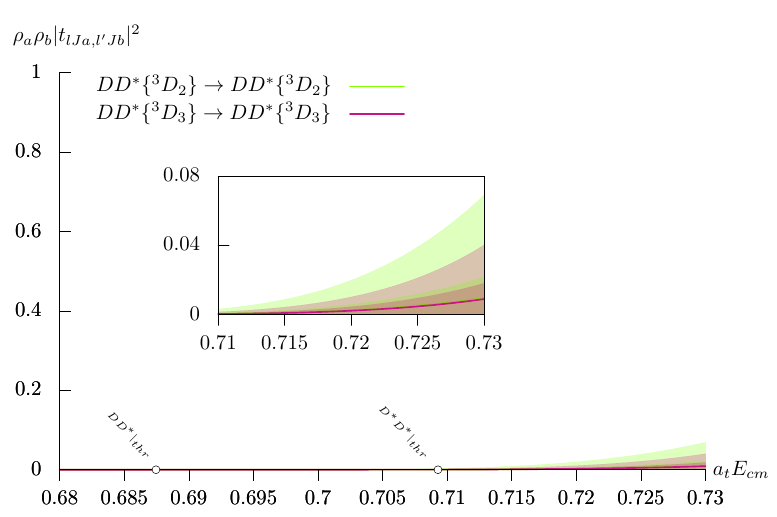}
\includegraphics[width=0.95\columnwidth]{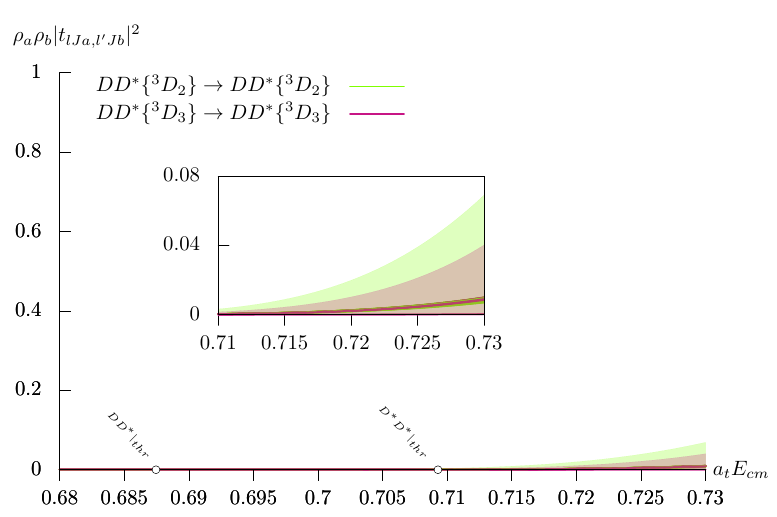}

\includegraphics[trim={0cm 0cm 0cm 6.5cm},clip,width=0.95\columnwidth]{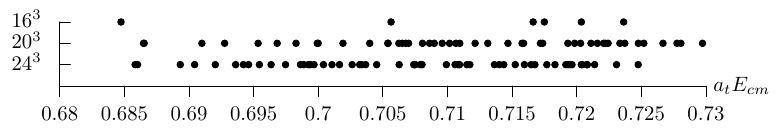}
\includegraphics[trim={0cm 0cm 0cm 6.5cm},clip,width=0.95\columnwidth]{coupled_channel_energies_scatter.pdf}
	\caption{\label{fig:coupled_channel_amps}(Left, top to bottom) Scattering amplitudes for \SLJ{3}{S}{1}, \SLJ{3}{D}{1}, \SLJ{3}{P}{0,2}, \SLJ{3}{D}{2,3} and \SLJ{3}{F}{2} for $\DDst$ and \SLJ{3}{S}{1}, \SLJ{3}{D}{1}, \SLJ{5}{P}{2} and \SLJ{5}{F}{2} for $\DstDst$ arising from the reference parameterization given by Eq.~(\ref{eq:ref_param}). Inner bands denote the statistical uncertainty while outer bands denote the systematic uncertainty obtained from varying the $D$ and $D^{\ast}$ masses and the anisotropy. (Right, top to bottom) Scattering amplitudes arising from all parameterizations given in Table \ref{tab:param_variations} with $\chi^2/N_{\mathrm{dof}} < 1.5$. The shaded band denotes the systematic uncertainty for the reference parameterization. (Bottom) The energy levels used in the scattering analysis.}
\end{figure*}
Figure \ref{fig:coupled_channel_amps} (left) shows the amplitudes arising from the reference parameterization of Eq.~(\ref{eq:ref_param}) and for all parameterization variations (right) given in Table \ref{tab:param_variations}. In all parameterizations, we observe a rapid opening of the $\DDst$ $S$-wave amplitude at threshold and a pronounced cusp-like feature peaked at $\DstDst$ threshold. The diagonal $\DstDst\rightarrow\DstDst$ and off-diagonal $\DDst\rightarrow \DstDst$ amplitudes both display a rapid opening at $\DstDst$ threshold.
These three features of the amplitudes suggest the presence of a pole close to the $\DstDst$ threshold, which would account for the cusp.
We also observe that the \SLJ{3}{D}{1} waves and the coupling between \SLJ{3}{S}{1} and \SLJ{3}{D}{1} waves in both hadron-hadron channels are consistent with zero. The other waves in $J^P = 0^-,2^-,2^+$ and $3^+$ are well constrained in the region near $\DstDst$ threshold with other waves having a partial wave configuration of $\ell \geq 2$ being statistically consistent with zero. As in the elastic case, we observe that the $\DDst$ amplitudes in the \SLJ{3}{P}{0} and \SLJ{3}{P}{2} channels are significantly nonzero, with a large difference in magnitude at higher energies.

To summarize our key findings of these various amplitude variations, as can be seen in Figs.~\ref{fig:amps_elastic} and~\ref{fig:coupled_channel_amps} for both the elastic and coupled-channel cases, we find a strong $S$-wave interaction, moderate $P$-wave interactions and only weak interactions in $D$-wave and higher. The coupled-channel effects are significant between $\DDst\SLJc{3}{S}{1}$ and $\DstDst\SLJc{3}{S}{1}$ with a pronounced cusp clearly visible. There is no appreciable coupling found between \SLJ{3}{S}{1} and \SLJ{3}{D}{1} in both hadron-hadron channels. The $\DDst\SLJc{3}{P}{0}$ is the next most significant amplitude - it shows no sharp features but is large across both elastic and coupled-channel scattering amplitudes. We now turn to investigating the singularities in the scattering amplitudes, from which masses, widths, and channel couplings of any states present can be obtained.

\clearpage
\section{\label{sec:poles}Scattering amplitude singularities}

Unstable hadrons can appear as resonating peaks in the scattering of lighter, stable hadrons. The least model-dependent way to describe these, that can unify bound states and resonances in multiple channels, is as the complex poles of a scattering amplitude. The real and imaginary parts of the pole position then correspond to the mass and width of the state, and the residue factorizes into the channel couplings. Sharp features in scattering amplitudes are often found to be due to a nearby pole. 

Experiment usually relies on production and decay processes, where an initial state excited by a high-energy interaction decays into the hadrons of interest. This is not simple $s$-channel scattering as has been computed here. However due to unitarity, the scattering $t$-matrix multiplies the production amplitude and thus the same singularities are present. 

The scattering amplitudes used in this work are analytic functions of $s=E_{\cm}^2$, except for branch cuts at each hadron-hadron threshold. These are thus relatively easy to search for complex poles, and such a procedure is well-defined. The amplitudes used do not include the full complexity that a complete description of these processes requires, such as the left-hand cuts arising below the $\DDst$ and $\DstDst$ thresholds. We will return to this issue later.

\subsection{\label{subsec:elastic_poles}Poles in $\DDst$ S-wave scattering}
We begin by considering the elastic $\DDst$ scattering region below $\DstDst$ threshold.
In our parameterization for elastic $\DDst$ scattering Eq.~(\ref{eq:elastic_params}), we found a good description of the finite volume energy levels with no $\SLJ{3}{S}{1}-\SLJ{3}{D}{1}$ coupling. In the limit that the $\SLJ{3}{D}{1}$ wave vanishes, then the existence of a pole in the elastic scattering $S$-wave $t$-matrix can be elucidated through the intersection of $a_t k \cot \delta_0$ with the curve $i a_t k$ at energies below threshold. If the $D$-wave amplitude is non-negligible in magnitude, the corrections to the $S$-wave phase shift from the determinant condition could significantly alter the intersection of $a_t k \cot \delta_0$. However, the $D$-wave amplitude was found to be small in Section \ref{sec:elastic} and such corrections are negligible to the $S$-wave phase shift and the intersection of $a_t k \cot \delta_0$ with $i a_t k$. This intersection can be seen in Fig.~\ref{fig:kcotdelta} and occurs at
\begin{equation}
	\mathrm{Im}(a_t k) = -0.0326(25)
\end{equation}
at an energy of
\begin{equation}
	(a_t E_{\cm}) = a_t \sqrt{s_0} = 0.6844(5),
\end{equation}
which corresponds to a virtual bound state. 

The couplings to the hadron channels are more straightforwardly determined from the $t$-matrix. At energies close to the position of a pole, the scattering $t$-matrix can be written
\begin{equation}
	t_{lSJa,l'S'Jb} \sim \frac{c_{lSJa}c_{l'S'Jb}}{s_0-s}
	\label{eq:fact_tmat}
\end{equation}
where the residue is factorized into the couplings $c_{lSJa}$ and $c_{l'S'Jb}$ that quantify the strength of the coupling to the decay channel. Performing this factorization for the $t$-matrix parameterized with the reference parameterization of Eq.~(\ref{eq:elastic_params}), we find the coupling of the virtual bound state to be
\begin{equation}
	a_t c^{\DDst\SLJc{3}{S}{1}} = 0.2571(69).
\end{equation}

We expand upon the position and coupling of this pole in the subsequent subsection using the amplitudes determined in coupled-channel analysis of Section \ref{sec:coupled}.

\subsection{\label{subsec:swave_poles}Poles in coupled-channel $\DDst-\DstDst$ S-wave scattering}
In coupled-channel $\DDst$-$\DstDst$ scattering, the complex $s$-plane is divided into 4 sheets due to the branch cut present at each threshold. The sheets are labelled by the sign of the imaginary part of the cm-frame momentum for each hadron channel using the notation $(\mathrm{sign}[\mathrm{Im}(k_{\DDst}]),\mathrm{sign}[\mathrm{Im}(k_{\DstDst})])$. The physical sheet, where physical scattering occurs, is defined by $\mathrm{Im}(k_{\DDst}), \mathrm{Im}(k_{\DstDst}) > 0$. We refer to this as sheet I, or equivalently the $(+,+)$ sheet. Causality forbids that poles off the real energy axis occur on the physical sheet. However, they are allowed to occur on the unphysical sheets II $(-,+)$, III $(-,-)$ and IV $(+,-)$ where they occur as complex-conjugate pairs in $s$. It is poles on these unphysical sheets that correspond to resonances with $\sqrt{s_0} = m \pm \frac{i}{2}\Gamma$, where $m$ is the mass of the resonance and $\Gamma$ its width. A pole off the real energy axis on an unphysical sheet can influence physical scattering if it is close to the physical sheet. In the case of sheets II and III, it is the lower half plane that is in close proximity, whereas for sheet IV it is the upper half. We will denote the relevant sheets using the shorthand $\mathrm{I}_u,\mathrm{II}_l,\mathrm{III}_l$ and $\mathrm{IV}_u$. Close to the pole, the scattering $t$-matrix is of the form of Eq.~(\ref{eq:fact_tmat}).

\begin{figure*}
\includegraphics[width=0.95\columnwidth]{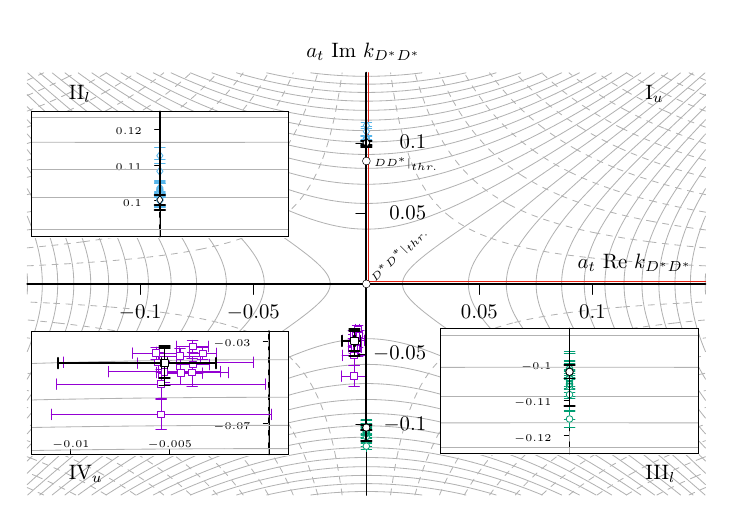}
\includegraphics[width=0.95\columnwidth]{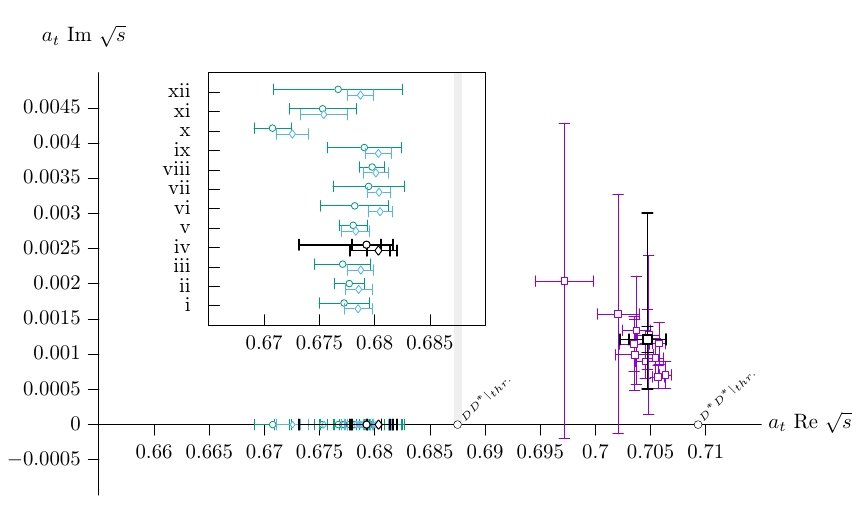}
	\caption{\label{fig:poles_mom_sqrts} (Left) The positions of the poles in the complex $\DstDst$ momentum plane for all parameterizations with a $\chi^2/N_{\mathrm{dof}} < 1.5$. Blue diamonds denote the position of the pole in sheet II, green circles denote the position of the pole in sheet III and purple squares the position of the pole in sheet IV. The positions of the poles arising from the reference parameterization are given in black, where the inner error bars are statistical and the outer error bars are the systematic error arising from varying the mass and anisotropy as described in Section \ref{sec:elastic}. Physical scattering is denoted by the red line in sheet I and grey contour lines are plotted for constant real (solid) and imaginary (dashed) $\sqrt{s}$. (Right) The pole positions in the complex $\sqrt{s}$ plane. The insert shows the position of the sheet II and sheet III poles for each parameterization, which have been displaced vertically for clarity.}
\end{figure*}
For the case of coupled-channel $\DDst-\DstDst$ scattering in $J^P=1^+$, three relevant poles are found for all parameterizations that adequately describe the finite volume spectra. Figure \ref{fig:poles_mom_sqrts} shows the position of the three poles in the complex $k_{\DstDst}$ plane (left) and their position in the complex $\sqrt{s}$-plane (right). A pole at real $s$ on sheet $\mathrm{II}_l$ is found below $\DDst$ threshold, corresponding to the virtual bound state discussed in the elastic analysis of Section \ref{sec:elastic}, along with a pole at real $s$ on sheet $\mathrm{III}_l$ below the $\DDst$ threshold. It is shown that the sheet $\mathrm{III}_l$ pole and $\mathrm{II}_l$ poles are related for the reference parameterization in the subsequent subsection. More interestingly, we observe a pole in $\mathrm{IV}_u$ with a very narrow width that lies close to the real axis and slightly below $\DstDst$ threshold. In \cite{Dudek:2016cru} it was shown that complex poles in $\mathrm{II}_l$ and $\mathrm{IV}_u$ can influence physical scattering, and induce an observable cusp in the amplitude. In Section \ref{subsec:coupling}, we will show that the $\mathrm{IV}_u$ pole is responsible for the cusp in the $\DDst$ $S$-wave amplitude.
\begin{figure*}[!htbp]
	\includegraphics[width=0.85\columnwidth]{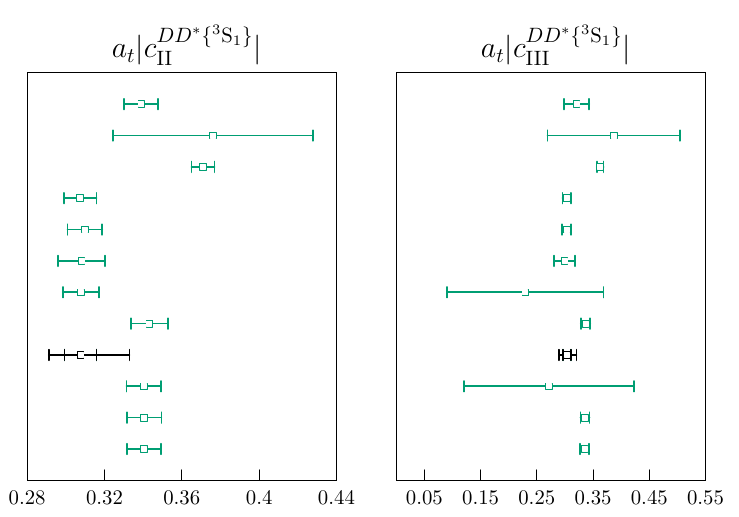}
	\includegraphics[width=0.85\columnwidth]{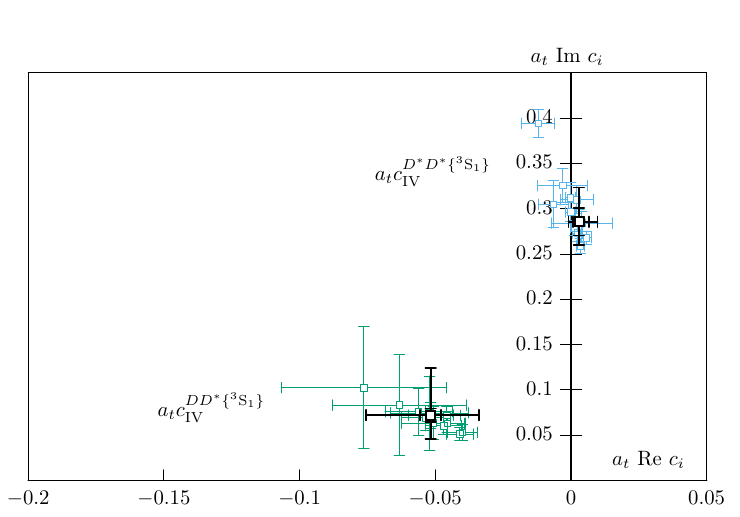}
	\caption{\label{fig:poles_coupling} The values of the coupling of the poles to $\DDst \SLJc{3}{S}{1}$ for the sheet II (left) and sheet III (center) poles and the couplings of the sheet IV pole to $\DDst \SLJc{3}{S}{1}$ and $\DstDst \SLJc{3}{S}{1}$ (right). The couplings from the reference parameterization are given in black, where the inner error bars are statistical  and the outer error bars are the systematic error from varying the mass and anisotropy as described in Section \ref{sec:elastic}.}
\end{figure*}

Taking a conservative envelope over the pole positions and couplings for all reasonable parameterizations, we find for the three poles
\begin{equation}
\begin{split}
	a_t \sqrt{s}_{\mathrm{II}} & = 0.6765(55)\\ 
	a_t \sqrt{s}_{\mathrm{III}} & = 0.6759(68)\\ 
	a_t \sqrt{s}_{\mathrm{IV}} &  = 0.7007(62) + \frac{i}{2}0.0020(22) \\
\end{split}
\nonumber
\end{equation}
with couplings
\begin{equation}
\begin{split}
	a_t c_{\mathrm{II}}^{{\DDst} \SLJc{3}{S}{1}} & = i0.36(7)\\ 
	a_t c_{\mathrm{III}}^{{\DDst} \SLJc{3}{S}{1}} & = i0.30(21)\\
a_t c_{\mathrm{IV}}^{{\DDst} \SLJc{3}{S}{1}} & = -0.07(4) + i0.10(7)\\ 
a_t c_{\mathrm{IV}}^{{\DstDst} \SLJc{3}{S}{1}} & = 0.00(4)+ i0.33(8)\\ 
\end{split}
\nonumber
\end{equation}
where the uncertainties are taken as the envelope of the individual uncertainties for each parameterization, and the central values are taken from the center of the envelope. The couplings of the poles for all parameterizations with $\chi^2/N_{\mathrm{dof}} < 1.5$ are given in Figure \ref{fig:poles_coupling}. The couplings of the sheet II and sheet III poles to $\DstDst$ \SLJc{3}{S}{1} are not reported due to the distance of these poles to the $\DstDst$ threshold. The sheet IV pole has nonzero couplings to both channels with the real part of the coupling to the $\DstDst$ channel being consistent with zero. In addition, the width is consistent with zero. All three poles have a small coupling to $D$-wave in both hadron-hadron channels, but due to the negligible $D$-wave amplitudes in both hadron-hadron channels, we do not report them.

Though the value of the width of the sheet IV pole shows a statistical consistency with zero using a conservative envelope over the pole positions from each parameterization, we show in the subsequent section for the reference parameterization that the coupling to both $\DDst \SLJc{3}{S}{1}$ and $\DstDst \SLJc{3}{S}{1}$ is necessarily non zero, as is the width.

\subsection{\label{subsec:coupling}Varying the $\DDst \rightarrow \DstDst$ $S$-wave Amplitude}
We examine a variation of the reference parameterization Eq.~(\ref{eq:ref_param}) where the coupling between the $\DDst$ and $\DstDst$ is reduced to zero. This is used to examine the sensitivity of the reference parameterization to the observed coupling and also to observe the behavior of the sheet IV pole as the $\DstDst$ channel is decoupled from $\DDst$. We change the parameter $\gamma^{(0)}_{\DDst\SLJc{3}{S}{1} \rightarrow \DstDst\SLJc{3}{S}{1}}$, which couples the $\DDst$ and $\DstDst$ channels in $S$-wave, by decreasing it towards zero in constant intervals while keeping all other parameters fixed. Table \ref{tab:mirror_pole} displays the values of $\gamma^{(0)}_{\DDst \SLJc{3}{S}{1} \rightarrow \DstDst \SLJc{3}{S}{1}}$ and the associated $\chi^2$ for the parameterization. We observe that as the value of the coupling parameter is decreased towards zero, the overall $\chi^2$ increases. 
This suggests that the determined spectra is incompatible with zero coupling between the channels for this parameterization.

\begin{figure}[!htbp]
\includegraphics[width=0.95\columnwidth]{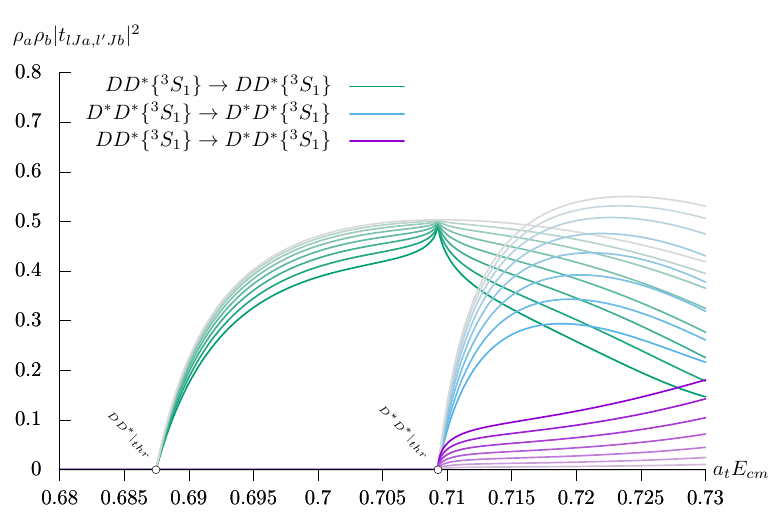}
\includegraphics[width=0.95\columnwidth]{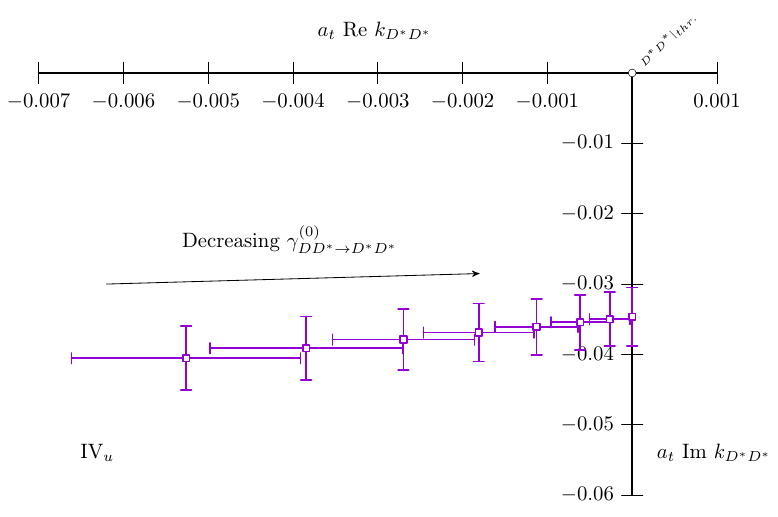}
\includegraphics[width=0.95\columnwidth]{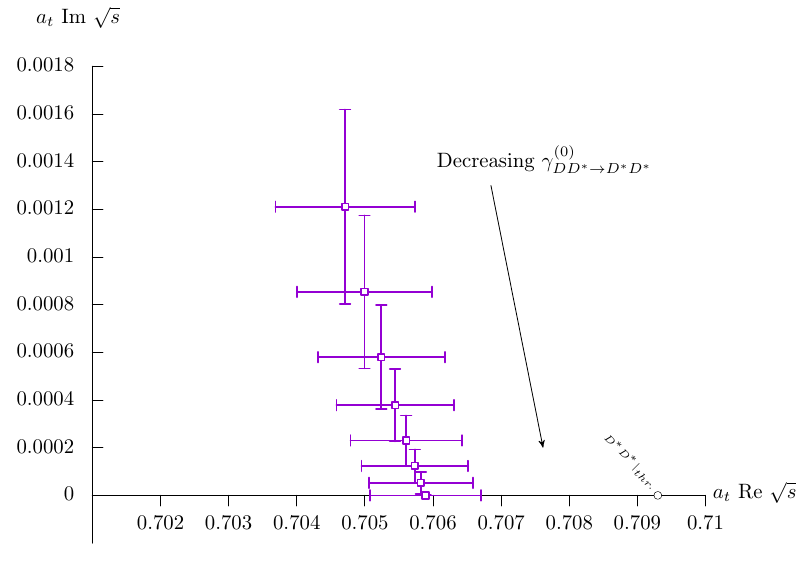}
\caption{\label{fig:sheetiv_traj}The scattering amplitude (top) and trajectory of the sheet IV pole in the complex $\DstDst$ momentum (middle) and complex energy (bottom) planes as the $\gamma^{(0)}_{\DDst \SLJc{3}{S}{1} \rightarrow \DstDst \SLJc{3}{S}{1}}$ matrix element goes to zero. The transition of the amplitudes from color to grey denotes a decrease in the value of the $\gamma^{(0)}_{\DDst \SLJc{3}{S}{1} \rightarrow \DstDst \SLJc{3}{S}{1}}$ matrix element.}
\end{figure}

Figure \ref{fig:sheetiv_traj} shows the amplitudes (top) and pole position (bottom) for each value of the coupling parameter. We observe that as the coupling parameter decreases, the near threshold cusp in the $\DDst$ $S$-wave amplitude becomes less pronounced, eventually disappearing when the coupling is completely turned off. This is accompanied by a rapid opening of the $\DstDst$ $S$-wave amplitude. The behavior of the amplitude can be explained by examining the position of the sheet IV pole. As the coupling parameter is decreased, the sheet IV pole moves to the axis with purely imaginary momentum in conjunction with the disappearance of the cusp. We interpret this resonance as a virtual bound state in the limit that the two hadron scattering channels are decoupled, which correspondingly enhances the $\DstDst$ $S$-wave amplitude upon opening. However, due to the open $\DDst$ channel and nonzero coupling, the pole is moved to a complex energy, manifesting as a resonance which is responsible for the cusp in the $\DDst$ $S$-wave amplitude. 

Also examined is the behavior of the sheet II and sheet III poles as the hadron-hadron channels are decoupled. Table \ref{tab:mirror_pole} shows the position of the sheet II and III poles in the complex $\DstDst$ momentum plane for the different values of the coupling parameter. When the two channels are fully decoupled, the sheet III pole becomes an exact ``mirror'' of the sheet II pole with opposite sign of momentum. Due to this behavior and in conjunction with its distance from the physical sheet, we do not prescribe it any physical significance.
\begin{table*}[!htbp]
\center
\begin{ruledtabular}
	\caption{\label{tab:mirror_pole}The $\chi^2$ of the fit and the positions of the sheet II and sheet III poles in the complex $\DstDst$ momentum plane as the $\DDst$-$\DstDst$ coupling parameter is reduced to zero. The \SLJ{3}{S}{1} label on \DDst ~and \DstDst ~has been suppressed for brevity in the parameter $\gamma$.}
\begin{tabular}{cccccc}
	& & \multicolumn{2}{c}{Sheet II} & \multicolumn{2}{c}{Sheet III} \\
	$\gamma^{(0)}_{\DDst \rightarrow \DstDst}$ & $\chi^2$ & Im$(k_{\DDst})$ & Im$(k_{\DstDst})$ & Im$(k_{\DDst})$ & Im$(k_{\DstDst})$ \\
\hline
	4.11 & 121.26 & -0.0493(36) & 0.1003(18) & -0.0529(43) & -0.1021(22) \\
	3.60 & 127.47 & -0.0480(35) & 0.0996(17) & -0.0507(39) & -0.1010(20) \\
	3.08 & 140.04 & -0.0469(34) & 0.0991(16) & -0.0489(37) & -0.1001(18) \\
	2.57 & 156.40 & -0.0460(32) & 0.0987(15) & -0.0473(33) & -0.0993(16) \\
	2.06 & 174.20 & -0.0453(31) & 0.0984(14) & -0.0461(31) & -0.0988(15) \\
	1.54 & 191.53 & -0.0448(31) & 0.0981(14) & -0.0452(31) & -0.0983(14) \\
	1.02 & 206.91 & -0.0444(30) & 0.0980(14) & -0.0446(30) & -0.0980(14) \\
	0 & 223.26 & -0.0441(32) & 0.0978(14) & -0.0441(32) & -0.0978(14) \\
\end{tabular}
\end{ruledtabular}
\end{table*}

To summarize our key findings on the coupling of the $\DDst$ and $\DstDst$ channels, we find that the coupling of the $\DDst \SLJc{3}{S}{1}$ and $\DstDst \SLJc{3}{S}{1}$ channels is necessarily nonzero in the reference parameterization in order to describe the finite volume spectra. The non zero width of the sheet IV pole is directly related to the coupling between the two channels and is responsible for the cusp observed in the $\DDst$ $S$-wave amplitude at the $\DstDst$ threshold energy. In addition, the sheet III pole is demonstrated to be a mirror pole of the sheet II pole. 

\subsection{\label{subsec:other_JP}Poles in other $J^P$}
The $1^+$ amplitudes contain $S$-waves, which are large and have sharp features. The majority of the other amplitudes determined are small and do not have features one usually associates with nearby poles. The next largest amplitude is $\SLJ{3}{P}{0}$, which rises relatively quickly for a $P$-wave and reaches a value $\rho^2|t|^2\approx 0.5$. We did not perform very many amplitude variations, given that there is not much constraint based on the energy levels present in the $[000]A^-_1$ and $A_2$ irreps. However, the reference parameterization in $J^P = 0^-$ contains a pole on the second Riemann sheet at
\begin{align}
a_t\sqrt{s}_\mathrm{II} &= (0.7011 \pm 0.0022) -\tfrac{i}{2}(0.062 \pm 0.011)\nonumber\\
a_t |c_{\DDst\SLJc{3}{P}{0}}| &= 0.2963 \pm 0.0125
\end{align}
where the quoted uncertainties are statistical. We stress that this is just a single parameterisation and so no firm conclusions should be drawn. With such a small amount of constraint, higher order amplitudes can easily introduce complex poles on the physical sheet\footnote{This occurs in all amplitudes attempted with a $K$-matrix with a linear polynomial in $s$.}. Given that relatively few energy levels are available, we do not investigate this further. However, it is possible this resonance pole is a feature in $J^P=0^-$ $\DDst\SLJc{3}{P}{0}$ scattering. The width is large, more than 300 MeV in physical units.

Other more distant poles can be found in some of the other partial waves. We do not consider any of these important as they are far away from where we constrain the scattering amplitudes and have a negligible effect on physical scattering.

\section{\label{sec:interp}Interpretation}

In all parameterizations capable of describing the finite volume spectra, we find a virtual bound state below $\DDst$ threshold which we identify with the $T_{cc}$, accompanied by a mirror pole on sheet III. In addition, we find a nearby complex resonance pole on sheet IV below $\DstDst$ threshold, which we denote $T_{cc}^\prime$, an as-yet unobserved state. This is responsible for the cusp in the $\DDst$ $S$-wave amplitude, and rapid rise in the $\DstDst$ $S$-wave amplitude. 

In order to present results in physical units, the lattice scale has been set using the $\Omega$ baryon mass as described in Section \ref{sec:lattice_setup}. The virtual bound state's position and coupling are found to be
\begin{equation}
\begin{split}
\sqrt{s} &= 3834(31)  \, \mathrm{MeV} \\
|c^{\DDst \SLJc{3}{S}{1}}| & = 2040(400) \, \mathrm{MeV,} \\
\end{split}
\nonumber
\end{equation}
where the uncertainties are taken as the envelope of the uncertainties for the pole position and coupling found from each parameterization, and the central values are taken to be the center of the envelope. The position of the virtual bound state is approximately $62$ MeV below $\DDst$ threshold.

\begin{figure*}
\begin{minipage}{0.95\columnwidth}
\includegraphics[width=0.95\columnwidth]{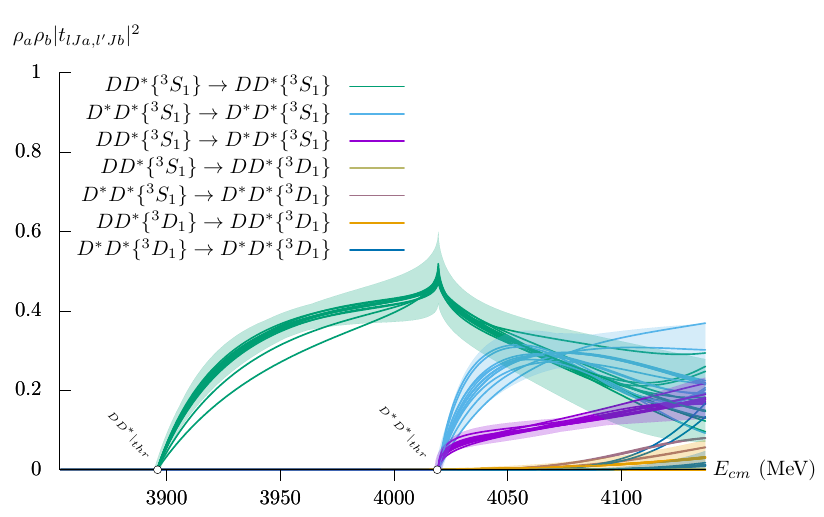}
\includegraphics[trim={0cm 0cm 0cm 6.5cm},clip,width=0.95\columnwidth]{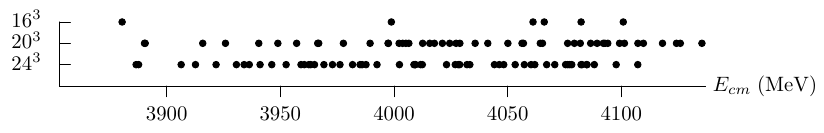}
\end{minipage}
\begin{minipage}{0.95\columnwidth}
\vspace{0.3cm}
\includegraphics[width=0.95\columnwidth]{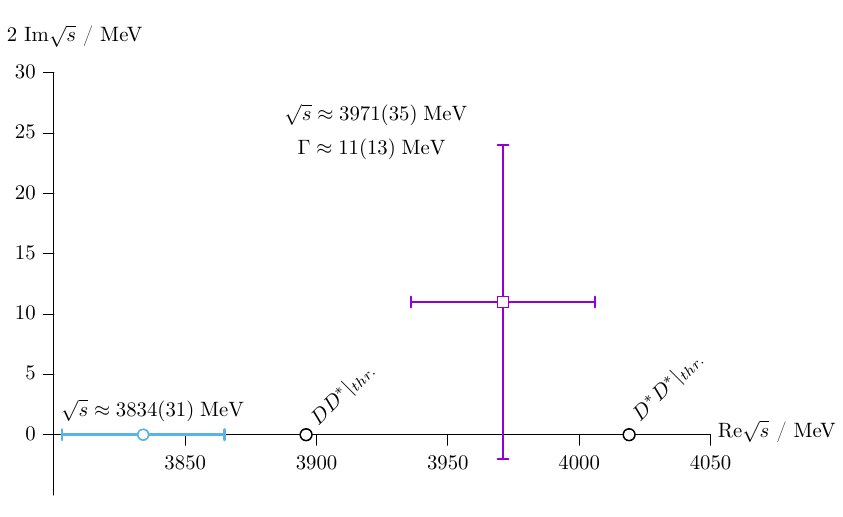}
\includegraphics[trim={0cm 0cm 0cm 6.5cm},clip,width=0.95\columnwidth]{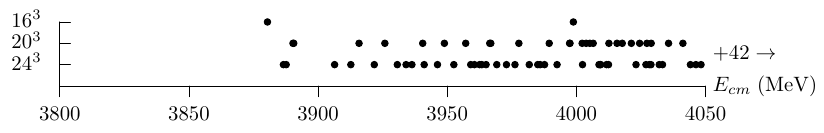}
\end{minipage}
	\caption{\label{fig:summary_fig}Left: The extracted $J^P = 1^+$ amplitudes for each parameterization with $\chi^2/N_{\mathrm{dof}} < 1.5$ (top) and the energy levels used to constrain them (bottom). The shaded band denotes the systematic uncertainty for the reference parameterization. Right: The positions of the sheet II virtual bound state and the sheet IV resonance in the complex energy plane (top) with the extracted energy levels (bottom).}
\end{figure*}

For the $T_{cc}^\prime$ resonance pole on sheet IV, the pole position and couplings are found to be
\begin{equation}
\begin{split}
\sqrt{s} &= 3971(35) + \frac{i}{2}11(13) \, \mathrm{MeV} \\
	|c^{\DstDst \SLJc{3}{S}{1}}| &=  1870(450) \, \mathrm{MeV} \\
	|c^{\DDst \SLJc{3}{S}{1}}| &=  692(195) \, \mathrm{MeV},\\
\end{split}
\nonumber
\end{equation}
which is approximately $49$ MeV below $\DstDst$ threshold, coupled most strongly to the kinematically closed $\DstDst$ channel. We also examined the coupling of the virtual bound state and the resonance to both hadron-hadron channels in $D$-wave, and found that the coupling of both poles is negligible.
The results are summarized in Fig. \ref{fig:summary_fig}, where we plot the $J^P = 1^+$ scattering amplitudes, the energy levels used to constrain them, and the positions of the virtual bound state and the resonance in the complex energy plane.

\subsection{\label{sec:systematics}Systematic Effects and Left-Cut Physics}

Working at a heavier than physical light-quark mass, that produces a pion with $m_\pi\approx 391$ MeV, a direct comparison with experiment is not possible. Since the experimentally-observed $T_{cc}$ is so close to threshold, fine details like the isospin splitting of the $D^0D^{\ast+}$ and $D^+D^{\ast0}$ thresholds and effects due to the unstable vector $D^\ast$ (which is stable in this calculation) may be significant in determining the properties of the $T_{cc}$.

With the scale setting used, the charm quarks produce an $\eta_c$ with $m_{\eta_c}\approx 2965$~MeV, slightly lower than the experimental value.
The presence of the heavy charm quark results in a relative insensitivity to the light-quark mass and so the $D$ and $D^\ast$ masses in this work are not that far from those in nature (the $D$ in this calculation has a mass 1886 MeV, and the $D^\ast$ mass is 2010 MeV).

In addition, the charm quarks used in this work produce a $J/\psi-\eta_c$ hyperfine splitting $\approx 33$~MeV lower than its experimental value~\cite{Liu:2012ze,Wilson:2023anv}. The hyperfine splitting of the $D^\ast$ and the $D$ is found to be 124 MeV, around 17~MeV lower than the observed hyperfine splittings (the charged and neutral cases differ slightly)~\cite{Moir:2013ub,Wilson:2023anv}. We thus expect systematic differences of a few 10s of MeV to be present. Notably, the $\DDst$ and $\DstDst$ thresholds are slightly closer to each other in this calculation than in nature.

The effects of left-cut physics arising from meson exchange in the scattering $t,u$-channels have not been accounted for in this calculation. They are also not currently included in the most general finite volume formalism applicable to coupled-channel scattering. The finite volume energy levels should include the necessary physics, but this has not been included in translating the lattice QCD energy levels into scattering information. At the masses used in this calculation, a $u$-channel cut due to $\pi$ exchange\footnote{This is the equivalent of the ``short nucleon cut'' cut seen in $N\pi$ scattering} ends at $a_t E_{cm}=0.68432$, roughly 18 MeV below threshold, and only a little below the lowest energy level extracted ($a_t E=0.68474(46)$ in $[000]T_1^+$ on the $L/a_s=16$ lattice). The virtual bound state pole we have identified would lie on this cut. The effect occurs again a similar distance below $\DstDst$ threshold at $a_t E_\cm = 0.70591$. Similarly to the virtual bound state, the resonance $T_{cc}^{\prime}$ identified is located at an energy close to the opening of this cut and it is possible that including the effects of left-cut physics in this calculation could change the properties of this resonance~\cite{Du:2023hlu}. Some modification is likely required to account for these cuts; at the very least, small imaginary parts must be introduced in the scattering amplitudes. Efforts are underway to add the necessary terms to the two-body L\"uscher formalism to include these effects~\cite{Raposo:2023nex,Raposo:2023oru}, though the formalism of these modifications is not yet mature enough to handle the coupled-channel case applicable to the $T_{cc}^\prime$. A second alternative for the $T_{cc}$ appears to be to continue the three-body formalism a short way below $\DDst$ threshold~\cite{Hansen:2024ffk}, however this is not applicable to the resonance $T_{cc}^\prime$, which would require four-body dynamics. As the amplitudes are consistent with the finite volume spectrum, it is possible that the effects of the left-cut physics are not that large.

Lippmann-Schwinger and effective field theory methods applied to the $\DDst$ system at larger-than-physical pion masses, where the $D^\ast$ is stable, can be found in Refs.~\cite{Du:2023hlu,Meng:2023bmz,Collins:2024sfi}. There it is shown that the presence of the left-cut below $\DDst$ threshold arising from one pion exchange affects both the nature of the $T_{cc}$ and its position. Similar complications can be expected below $\DstDst$ threshold for the $T_{cc}^\prime$, and so extra work is needed to understand how the structures observed in this study are affected.

\subsection{\label{sec:other_work}Comparison to Previous Work and Phenomenology}

Other works have studied this channel. Considering only elastic scattering, Ref. \cite{Padmanath:2022cvl} working at a light-quark mass corresponding to $m_\pi=280$~MeV, finds a virtual bound state 9.9 MeV below $\DDst$ threshold in $S$-wave. Using the HAL QCD method with almost-physical pions, Ref.~\cite{Lyu:2023xro} finds a virtual bound state much closer to threshold. While these results point toward a ground-state $T_{cc}$ that gets closer to threshold with decreasing light-quark mass, further work is needed to understand the structures in the complex plane before firm conclusions can be made. In Ref. \cite{Cheung:2017tnt}, the spectrum of the $I = 0$ doubly charmed sector was computed and no strong indications of a bound state or narrow resonance were present. However, the authors noted that ``the situation is not as straightforward when one considers coupled-channel scattering or broad resonances". The identification of both the $T_{cc}$ and the $T_{cc}^{\prime}$ highlights the importance of extracting enough energy levels to perform a scattering analysis in order to elucidate the pole structure in coupled-channel systems.

Phenomenological studies have previously suggested that strong effects may be seen at $\DstDst$ threshold~\cite{Molina:2010tx,Li:2012ss,Sun:2012zzd,Liu:2019stu,Dong:2021bvy,Dai:2021vgf,Ortega:2022efc}. One source of motivation for this is the infinitely heavy quark limit where the $D$ and $D^\ast$ become degenerate, and the interactions of $\DDst\SLJc{3}{S}{1}$ and $\DstDst\SLJc{3}{S}{1}$ are identical~\cite{Albaladejo:2021vln}. Away from this limit, the degeneracy is broken and the relation no longer holds. However, given the large charm quark mass relative to the other relevant energy scales, corrections which should be proportional to $1/m_c$ could be modest. Similarities are found between $\DDst$ and $\DstDst$ close to threshold in $\SLJ{3}{S}{1}$ in this calculation. Phenomenological approaches do not usually quantify the mixing between $\DDst\SLJc{3}{S}{1}$ and $\DstDst\SLJc{3}{S}{1}$; it is often assumed to be zero. Here we have found it to be non-zero and significant. 

In experiment, the $T_{cc}^+(3875)$ and $X/\chi_{c1}(3872)$ are seen close to thresholds involving a vector $D^\ast$-meson. These produce apparent excesses in $DD$ or $D\bar{D}$ final states respectively, due to the vector $D^\ast$ decaying to a $D$ and a low-momentum pion, or photon, that then escape detection. This is explained in Ref.~\cite{LHCb:2021auc}, see for example Fig.~7 therein, where peaks are visible close to the pseudoscalar-pseudoscalar thresholds. With the $T_{cc}^\prime$ in mind, the present work suggests it should be visible in both $\DDst$ and $\DstDst$ final states. One may then anticipate that a contribution from $T_{cc}^\prime$ might already be present in the data and could possibly be seen in $DD$ and/or $DD\pi$ mass distributions. There are no other clear peaks visible in Ref.~\cite{LHCb:2021auc}, but it is possible that they may be revealed by future analyses with improved data.

In this work we also computed several partial waves that were expected to be suppressed relative to the \SLJ{3}{S}{1} amplitudes. In particular, we found $\DDst\SLJc{3}{P}{0}$ to be relatively large.  Large, statistically-significant downward shifts are seen in $[000]A_1^-$ where this partial wave appears in relative isolation. The resulting amplitudes describing the levels were large enough to support resonance poles in some cases, although we also found descriptions with very different poles, and left-cut effects are also present~\cite{Meng:2023bmz}. Nevertheless, the interactions in $\DDst\SLJc{3}{P}{0}$ appear quite strongly attractive.

The magnitude of the $P$-wave amplitudes in $J^P = 0^-$ and $2^-$ are significantly different. An argument for the observed asymmetry comes from the one pion exchange model. In this model, the potential is directly proportional to the spin matrix element in the tensor interaction \cite{Tornqvist:1993ng,Thomas:2008ja}. The magnitude of this matrix element is directly proportional to the Racah coefficient, which for $J^P = 0^-$ and $2^-$ is $1/3$ and $1/30$, respectively. Such a difference in magnitude for the spin matrix element leads to a weaker potential for $J^P = 2^-$ than for $0^-$, leading to a smaller scattering amplitude as is observed in this calculation.

\section{\label{sec:summary}Summary}
In summary, we have presented the first coupled-channel $\DDst$, $\DstDst$ scattering calculation from lattice QCD in $S$-wave for isospin-0. Using 109 energy levels, we determined the $J^P = 1^+$ scattering amplitude by constraining other waves in the $\DDst$ and $\DstDst$ channels for 12 parameterization variations that adequately describe the finite volume spectrum. In all successful variations, we observe a pole in the $S$-wave amplitudes below $\DDst$ threshold corresponding to a virtual bound state at approximately $\sqrt{s} \approx 3834(31)$ MeV, commonly identified as the $T_{cc}$. A pole corresponding to a resonance $T_{cc}^\prime$ is also found just below $\DstDst$ threshold at $\sqrt{s} \approx 3971(35)$ MeV with a width of $\Gamma \approx 11(13)$ MeV, which predominantly couples to the kinematically closed $\DstDst$ channel. The presence of the nearby resonance pole below $\DstDst$ threshold enhances the $\DDst$ $S$-wave amplitude, resulting in clear cusp at $\DstDst$ threshold. As ongoing experiments are currently investigating the isospin-0 $J^P = 1^+$ charm sector, it should be possible to find the experimental $T_{cc}^\prime$ in both $\DDst$ and $\DstDst$ final states.

\begin{acknowledgments}
	We thank our colleagues within the Hadron Spectrum Collaboration (\url{www.hadspec.org}), particularly Jozef Dudek, Raul Brice\~{n}o and  Andrew Jackura for useful discussions. We would also like to thank Mikhail Mikhasenko for many insightful discussions. 
	TW \& DJW acknowledge support from a Royal Society University Research Fellowship. TW, DJW \& CET acknowledge support from the U.K. Science and Technology Facilities Council (STFC) [grant numbers ST/T000694/1, ST/X000664/1]. TW also acknowledges support from Science Foundation Ireland [grant number 21/FFP-P/10186]

The software codes
{\tt Chroma}~\cite{Edwards:2004sx}, {\tt QUDA}~\cite{Clark:2009wm,Babich:2010mu}, {\tt QUDA-MG}~\cite{Clark:SC2016}, {\tt QPhiX}~\cite{ISC13Phi}, {\tt MG\_PROTO}~\cite{MGProtoDownload}, {\tt QOPQDP}~\cite{Osborn:2010mb,Babich:2010qb}, and {\tt Redstar}~\cite{Chen:2023zyy} were used. 
Some software codes used in this project were developed with support from the U.S.\ Department of Energy, Office of Science, Office of Advanced Scientific Computing Research and Office of Nuclear Physics, Scientific Discovery through Advanced Computing (SciDAC) program; also acknowledged is support from the Exascale Computing Project (17-SC-20-SC), a collaborative effort of the U.S.\ Department of Energy Office of Science and the National Nuclear Security Administration.

This work used the Cambridge Service for Data Driven Discovery (CSD3), part of which is operated by the University of Cambridge Research Computing Service (www.csd3.cam.ac.uk) on behalf of the STFC DiRAC HPC Facility (www.dirac.ac.uk). The DiRAC component of CSD3 was funded by BEIS capital funding via STFC capital grants ST/P002307/1 and ST/R002452/1 and STFC operations grant ST/R00689X/1. Other components were provided by Dell EMC and Intel using Tier-2 funding from the Engineering and Physical Sciences Research Council (capital grant EP/P020259/1). This work also used the earlier DiRAC Data Analytic system at the University of Cambridge. This equipment was funded by BIS National E-infrastructure capital grant (ST/K001590/1), STFC capital grants ST/H008861/1 and ST/H00887X/1, and STFC DiRAC Operations grant ST/K00333X/1. DiRAC is part of the National E-Infrastructure.
This work also used clusters at Jefferson Laboratory under the USQCD Initiative and the LQCD ARRA project.

Propagators and gauge configurations used in this project were generated using DiRAC facilities, at Jefferson Lab, and on the Wilkes GPU cluster at the University of Cambridge High Performance Computing Service, provided by Dell Inc., NVIDIA and Mellanox, and part funded by STFC with industrial sponsorship from Rolls Royce and Mitsubishi Heavy Industries. Also used was an award of computer time provided by the U.S.\ Department of Energy INCITE program and supported in part under an ALCC award, and resources at: the Oak Ridge Leadership Computing Facility, which is a DOE Office of Science User Facility supported under Contract DE-AC05-00OR22725; the National Energy Research Scientific Computing Center (NERSC), a U.S.\ Department of Energy Office of Science User Facility located at Lawrence Berkeley National Laboratory, operated under Contract No. DE-AC02-05CH11231; the Texas Advanced Computing Center (TACC) at The University of Texas at Austin; the Extreme Science and Engineering Discovery Environment (XSEDE), which is supported by National Science Foundation Grant No. ACI-1548562; and part of the Blue Waters sustained-petascale computing project, which is supported by the National Science Foundation (awards OCI-0725070 and ACI-1238993) and the state of Illinois. Blue Waters is a joint effort of the University of Illinois at Urbana-Champaign and its National Center for Supercomputing Applications.
\end{acknowledgments}

\section*{Data Availability}

Reasonable requests for data, such as energy levels and correlations, can be directed to the authors and will be considered in accordance with the Hadron Spectrum Collaboration's policy on sharing data.

\clearpage
\pagebreak

\appendix

\onecolumngrid

\section{\label{sec:sub_tables}Subduction Tables}
In this Appendix, we list the pseudoscalar-vector and identical vector-vector partial waves that contribute to the finite volume irreducible representations in Tables \ref{tab:pv_atrest}-\ref{tab:vv_inflight}. Note that for identical vector-vector scattering, certain partial waves are not allowed by Bose symmetry (see Table \ref{tab:DstDstPWs}) and these are ommitted from the tables.
\begin{table*}[!h]
{\renewcommand{\arraystretch}{1.2}
\begin{tabular}{c|c|c|c|c|c}
$\Lambda^{P}$ & $T^+_1$ & $A^-_1$ & $E^-$ & $E^+$ & $A^+_2$ \\
\hline
&&&&&\\[-2.5ex]
	\multirow{5}{5em}{$J^P(\SLJ{2S+1}{\ell}{J})~$} &  & $0^- \left( \SLJ{3}{P}{0}  \right)$ & & & \\
& $1^+ \left( \begin{matrix} \SLJ{3}{S}{1} \\ \SLJ{3}{D}{1} \end{matrix} \right)$ & & & \\
& & & $2^- \left( \begin{matrix} \SLJ{3}{P}{2} \\ \SLJ{3}{F}{2} \end{matrix} \right)$ & $2^+ \left( \SLJ{3}{D}{2}  \right)$ & \\
& $3^+ \left( \begin{matrix} \SLJ{3}{D}{3} \\ \SLJ{3}{G}{3} \end{matrix} \right)$ & & & & $3^+ \left( \begin{matrix} \SLJ{3}{D}{3} \\ \SLJ{3}{G}{3} \end{matrix} \right)$ \\
	& $4^+ \left( \SLJ{3}{G}{4}  \right)$  & $4^- \left( \begin{matrix} \SLJ{3}{F}{4} \\ \SLJ{3}{H}{4} \end{matrix} \right)$ & $4^- \left( \begin{matrix} \SLJ{3}{F}{4} \\ \SLJ{3}{H}{4} \end{matrix} \right)$ & $4^+ \left( \SLJ{3}{G}{4}  \right)$ & \\ [2.5ex]
\hline
\end{tabular}
\caption{\label{tab:pv_atrest}Subduction table for pseudoscalar-vector scattering overall at rest.}
}
\end{table*}

\begin{table*}[!h]
{\renewcommand{\arraystretch}{1.2}
\begin{tabular}{c|c|c|c|c|c}
$\Lambda^{P}$ & $T^+_1$ & $A^-_1$ & $E^-$ & $E^+$ & $A^+_2$ \\
\hline
&&&&&\\[-2.5ex]
\multirow{5}{5em}{$J^P(\SLJ{2S+1}{\ell}{J})~$} & & & & & \\
& $1^+ \left( \begin{matrix} \SLJ{3}{S}{1} \\ \SLJ{3}{D}{1} \end{matrix} \right)$ & & & \\
& & & $2^- \left( \begin{matrix} \SLJ{5}{P}{2} \\ \SLJ{5}{F}{2} \end{matrix} \right)$ & $2^+ \left( \SLJ{3}{D}{2}  \right)$ & \\
& $3^+ \left( \begin{matrix} \SLJ{3}{D}{3} \\ \SLJ{3}{G}{3} \end{matrix} \right)$ & & & & $3^+ \left( \begin{matrix} \SLJ{3}{D}{3} \\ \SLJ{3}{G}{3} \end{matrix} \right)$ \\
	& $4^+ \left( \SLJ{3}{G}{4}  \right)$  & $4^- \left( \begin{matrix} \SLJ{5}{F}{4} \\ \SLJ{5}{H}{4} \end{matrix} \right)$ & $4^- \left( \begin{matrix} \SLJ{5}{F}{4} \\ \SLJ{5}{H}{4} \end{matrix} \right)$ & $4^+ \left( \SLJ{3}{G}{4}  \right)$ & \\ [2.5ex]
\hline
\end{tabular}
\caption{\label{tab:vv_atrest}Subduction table for identical vector-vector scattering overall at rest.}
}
\end{table*}

\begin{table*}[!h]
{\renewcommand{\arraystretch}{1.2}
\begin{tabular}{c|c|c|c}
$\Lambda^{P}$ & $[00n]A_2$ & $[0nn]A_2$ & $[nnn]A_2$ \\
\hline
&&&\\[-2.5ex]
\multirow{9}{5em}{$J^P(\SLJ{2S+1}{\ell}{J})~$} & $0^- \left( \SLJ{3}{P}{0}  \right)$ & $0^- \left( \SLJ{3}{P}{0}  \right)$  & $0^- \left( \SLJ{3}{P}{0}  \right)$ \\
&&&\\[-2.5ex]
& $1^+ \left( \begin{matrix} \SLJ{3}{S}{1} \\ \SLJ{3}{D}{1} \end{matrix} \right)$ & $1^+ \left( \begin{matrix} \SLJ{3}{S}{1} \\ \SLJ{3}{D}{1} \end{matrix} \right)$ &  $1^+ \left( \begin{matrix} \SLJ{3}{S}{1} \\ \SLJ{3}{D}{1} \end{matrix} \right)$\\
&&&\\[-2.5ex]
	& $\begin{matrix} \\ 2^- \left( \begin{matrix} \SLJ{3}{P}{2} \\ \SLJ{3}{F}{2} \end{matrix} \right) \end{matrix} $  & $\begin{matrix}2^+ \left( \SLJ{3}{D}{2}  \right) \\ 2^- \left( \begin{matrix} \SLJ{3}{P}{2} \\ \SLJ{3}{F}{2} \end{matrix} \right)_{[2]} \end{matrix} $ & $\begin{matrix} \\ 2^- \left( \begin{matrix} \SLJ{3}{P}{2} \\ \SLJ{3}{F}{2} \end{matrix} \right) \end{matrix} $ \\
&&&\\[-2.5ex]
	& $\begin{matrix} 3^+ \left( \begin{matrix} \SLJ{3}{D}{3} \\ \SLJ{3}{G}{3} \end{matrix} \right) \\ \phantom{3^- \left( \SLJ{3}{F}{3} \right)} \end{matrix}$ & $\begin{matrix} 3^+ \left( \begin{matrix} \SLJ{3}{D}{3} \\ \SLJ{3}{G}{3} \end{matrix} \right)_{[2]} \\ 3^- \left( \SLJ{3}{F}{3} \right) \end{matrix}$ & $\begin{matrix} 3^+ \left( \begin{matrix} \SLJ{3}{D}{3} \\ \SLJ{3}{G}{3} \end{matrix} \right)_{[2]} \\ 3^- \left( \SLJ{3}{F}{3} \right) \end{matrix}$\\
&&&\\[-2.5ex]
	& $4^- \left( \begin{matrix} \SLJ{3}{F}{4} \\ \SLJ{3}{H}{4} \end{matrix} \right)_{[2]}$  & $4^- \left( \begin{matrix} \SLJ{3}{F}{4} \\ \SLJ{3}{H}{4} \end{matrix} \right)_{[3]}$ & $4^- \left( \begin{matrix} \SLJ{3}{F}{4} \\ \SLJ{3}{H}{4} \end{matrix} \right)_{[2]}$ \\ [2.5ex]
\hline
\end{tabular}
\caption{\label{tab:pv_inflight}Subduction table for pseudoscalar-vector at overall nonzero momentum. The numbers in square brackets indicate the number of times a given amplitude appears (number of embeddings).}
}
\end{table*}

\begin{table*}[!h]
{\renewcommand{\arraystretch}{1.2}
\begin{tabular}{c|c|c|c}
$\Lambda^{P}$ & $[00n]A_2$ & $[0nn]A_2$ & $[nnn]A_2$ \\
\hline
&&&\\[-2.5ex]
\multirow{9}{5em}{$J^P(\SLJ{2S+1}{\ell}{J})~$} & & & \\
&&&\\[-2.5ex]
& $1^+ \left( \begin{matrix} \SLJ{3}{S}{1} \\ \SLJ{3}{D}{1} \end{matrix} \right)$ & $1^+ \left( \begin{matrix} \SLJ{3}{S}{1} \\ \SLJ{3}{D}{1} \end{matrix} \right)$ &  $1^+ \left( \begin{matrix} \SLJ{3}{S}{1} \\ \SLJ{3}{D}{1} \end{matrix} \right)$\\
&&&\\[-2.5ex]
	& $\begin{matrix} \\ 2^- \left( \begin{matrix} \SLJ{5}{P}{2} \\ \SLJ{5}{F}{2} \end{matrix} \right) \end{matrix} $  & $\begin{matrix}2^+ \left( \SLJ{3}{D}{2}  \right) \\ 2^- \left( \begin{matrix} \SLJ{5}{P}{2} \\ \SLJ{5}{F}{2} \end{matrix} \right)_{[2]} \end{matrix} $ & $\begin{matrix} \\ 2^- \left( \begin{matrix} \SLJ{5}{P}{2} \\ \SLJ{5}{F}{2} \end{matrix} \right) \end{matrix} $ \\
&&&\\[-2.5ex]
	& $\begin{matrix} 3^+ \left( \begin{matrix} \SLJ{3}{D}{3} \\ \SLJ{3}{G}{3} \end{matrix} \right) \\ \phantom{3^-  \left( \begin{matrix} \SLJ{5}{P}{3} \\ \SLJ{5}{F}{3}  \end{matrix} \right)} \end{matrix}$ & $\begin{matrix} 3^+ \left( \begin{matrix} \SLJ{3}{D}{3} \\ \SLJ{3}{G}{3} \end{matrix} \right)_{[2]} \\ 3^-  \left( \begin{matrix} \SLJ{5}{P}{3} \\ \SLJ{5}{F}{3}  \end{matrix} \right) \end{matrix}$ & $\begin{matrix} 3^+ \left( \begin{matrix} \SLJ{3}{D}{3} \\ \SLJ{3}{G}{3} \end{matrix} \right)_{[2]} \\ 3^-  \left( \begin{matrix} \SLJ{5}{P}{3} \\ \SLJ{5}{F}{3}  \end{matrix} \right) \end{matrix}$ \\
&&&\\[-2.5ex]
	& $4^- \left( \begin{matrix} \SLJ{5}{F}{4} \\ \SLJ{5}{H}{4} \end{matrix} \right)_{[2]}$  & $4^- \left( \begin{matrix} \SLJ{5}{F}{4} \\ \SLJ{5}{H}{4} \end{matrix} \right)_{[3]}$ & $4^- \left( \begin{matrix} \SLJ{5}{F}{4} \\ \SLJ{5}{H}{4} \end{matrix} \right)_{[2]}$ \\ [2.5ex]
\hline
\end{tabular}
\caption{\label{tab:vv_inflight}Subduction table for identical vector-vector mesons at overall nonzero momentum. The numbers in square brackets indicate the number of times a given amplitude appears (number of embeddings).}
}
\end{table*}

\clearpage
\pagebreak

\section{\label{sec:op_tables}Operator Tables}
In Tables \ref{tab:rest_ops} and \ref{tab:inflight_ops} we list the $\DDst$ and $\DstDst$ operators used in this study, with their relative momentum denoted by $[n_xn_yn_z]$, including their multiplicity. Operators listed in grey were included in the basis in the variational extraction of energy levels, but energy levels associated with these operators were not used in the scattering analysis. 
\begin{table*}[!h]
\begin{small}
\begin{tabular}{c|c|c|c|c|c}
\multicolumn{3}{c}{$[000]T_1^+$} &\multicolumn{3}{c}{$[000]A^-_1$}\\
\hline
$L/a_s=16$ & $L/a_s=20$ & $L/a_s=24$ & $L/a_s=16$ & $L/a_s=20$ & $L/a_s=24$ \\
\hline
$D[000] D^\ast[000]$ & $D[000] D^\ast[000]$ & $D[000] D^\ast[000]$ & $D[001] D^\ast[001]$ & $D[001] D^\ast[001]$ & $D[001] D^\ast[001]$ \\
$D^\ast[000] D^\ast[000]$ & $D^\ast[000] D^\ast[000]$ & $D[001] D^\ast[001] \times 2$ & \textcolor{lightgray}{$D[011] D^\ast[011]$} & \textcolor{lightgray}{$D[011] D^\ast[011]$} & $D[011] D^\ast[011]$ \\
$D[001] D^\ast[001]\times 2$ & $D[001] D^\ast[001]$  & $D^\ast[000] D^\ast[000]$ &  & \textcolor{lightgray}{$D^\ast[011] D^\ast[011]$} & \textcolor{lightgray}{$D[111] D^\ast[111]$}\\
$\textcolor{lightgray}{D^\ast[001] D^\ast[001]\times 2}$ & $D^\ast[001] D^\ast[001]\times 2$ & $D[011] D^\ast[011]\times 3$ &         & \textcolor{lightgray}{$D[111] D^\ast[111]$} & \textcolor{lightgray}{$D^\ast[011] D^\ast[011]$}  \\
$\textcolor{lightgray}{D[011] D^\ast[011]\times 3}$ & \textcolor{lightgray}{$D[011] D^\ast[011]\times 3$} & $D^\ast[001] D^\ast[001]\times 2$ &         &  & \textcolor{lightgray}{$D[002] D^\ast[002]$} \\
        & \textcolor{lightgray}{$D^\ast[011] D^\ast[011]\times 3$} & \textcolor{lightgray}{$D[111] D^\ast[111]\times 2$} &         &         &  \\
        & \textcolor{lightgray}{$D[111] D^\ast[111]\times 2$} & \textcolor{lightgray}{$D^\ast[011] D^\ast[011]\times 3$} &         &         &  \\
        &         & \textcolor{lightgray}{$D[002] D^\ast[002] \times 2$} & & & \\
        &         & \textcolor{lightgray}{$D^\ast[111] D^\ast[111] \times 2$} & & & \\
\multicolumn{6}{c}{} \\
\multicolumn{3}{c}{$[000]E^-$} &\multicolumn{3}{c}{$[000]E^+$} \\
\hline
$L/a_s=16$ & $L/a_s=20$ & $L/a_s=24$ & $L/a_s=16$ & $L/a_s=20$ & $L/a_s=24$ \\
\hline
$D[001] D^\ast[001]$ & $D[001] D^\ast[001]$ & $D[001] D^\ast[001]$ & \textcolor{lightgray}{$D[011] D^\ast[011]$} & \textcolor{lightgray}{$D[011] D^\ast[011]$} & $D[011] D^\ast[011]$  \\
\textcolor{lightgray}{$D[011] D^\ast[011]\times 2$} & $D^\ast[001] D^\ast[001]$ & $D[011] D^\ast[011]\times 2$ &         & \textcolor{lightgray}{$D[111] D^\ast[111]$} & \textcolor{lightgray}{$D[111] D^\ast[111]$}  \\
\textcolor{lightgray}{$D^\ast[001] D^\ast[001]$} & \textcolor{lightgray}{$D[011] D^\ast[011]\times 2$} & $D^\ast[001] D^\ast[001]$ &         & \textcolor{lightgray}{$D^\ast[011] D^\ast[011]$} & \textcolor{lightgray}{$D^\ast[011] D^\ast[011]$} \\
        & \textcolor{lightgray}{$D^\ast[011] D^\ast[011]\times 2$} & \textcolor{lightgray}{$D[111] D^\ast[111]$} &         &         & \textcolor{lightgray}{$D^\ast[111] D^\ast[111]$} \\
        &         & \textcolor{lightgray}{$D^\ast[011] D^\ast[011]\times 2$} &         &         &  \\
        &         & \textcolor{lightgray}{$D[002] D^\ast[002]$} &         &         &  \\
        &         & \textcolor{lightgray}{$D^\ast[111] D^\ast[111]$} & & \\
\multicolumn{6}{c}{} 
\end{tabular}
\begin{tabular}{c}
$[000]A^+_2$ \\
\hline
\begin{tabular}{c|c|c}
$L/a_s=16$ & $L/a_s=20$ & $L/a_s=24$\\
\hline
\textcolor{lightgray}{$D[011] D^\ast[011]$} & \textcolor{lightgray}{$D[011] D^\ast[011]$}   & $D[011] D^\ast[011]$ \\
        &\textcolor{lightgray}{$D^\ast[011] D^\ast[011]$} &\textcolor{lightgray}{$D^\ast[001] D^\ast[001]$} \\
        & \textcolor{lightgray}{$D[111] D^\ast[111]$}  & \textcolor{lightgray}{$D[111] D^\ast[111]$} \\
        & & \textcolor{lightgray}{$D^\ast[011] D^\ast[011]$} \\
        &         & \textcolor{lightgray}{$D^\ast[111] D^\ast[111]$} \\
\end{tabular}
\end{tabular}
\end{small}
\caption{\label{tab:rest_ops}Operators for the at rest irreps.}
\end{table*}

\begin{table*}[!h]
\begin{small}
\begin{tabular}{c|c|c|c}
\multicolumn{2}{c}{$[001]A_2$} & \multicolumn{2}{c}{$[002]A_2$}\\
\hline
$L/a_s=20$ & $L/a_s=24$ & $L/a_s=20$ & $L/a_s=24$ \\
\hline
$D[000] D^\ast[001]$ & $D[000] D^\ast[001]$ & $D[001] D^\ast[001]$ & $D[001] D^\ast[001]$ \\
$D[001] D^\ast[000]$ & $D[001] D^\ast[000]$ & $D[000] D^\ast[002]$ & $D[000] D^\ast[002]$ \\
$D^\ast[001] D^\ast[000]$ & $D[001] D^\ast[011]\times 2$ & $D^\ast[001] D^\ast[001]$ & $D[011] D^\ast[011]\times 2$ \\
$D[001] D^\ast[011]\times 2$ & $D[011] D^\ast[001]\times 2$ & $D[002] D^\ast[000]$ & $D[002] D^\ast[000]$ \\
$D[011] D^\ast[001]\times 2$ & $D^\ast[001] D^\ast[000]$ & $D[011] D^\ast[011]\times 2$ & $D^\ast[001] D^\ast[001]$ \\
\textcolor{lightgray}{$D^\ast[011] D^\ast[001]\times 4$} & \textcolor{lightgray}{$D[001] D^\ast[002]$} &  \textcolor{lightgray}{$D^\ast[002] D^\ast[000]$} &  \textcolor{lightgray}{$D[001] D^\ast[012]\times 2$} \\
\textcolor{lightgray}{$D[001] D^\ast[002]$} & \textcolor{lightgray}{$D[011] D^\ast[111]\times 2$} &  \textcolor{lightgray}{$D[001] D^\ast[012]\times 2$} &  \textcolor{lightgray}{$D[111] D^\ast[111]\times 2$} \\
\textcolor{lightgray}{$D[011] D^\ast[111]\times 2$} & \textcolor{lightgray}{$D[111] D^\ast[011]\times 2$} &  \textcolor{lightgray}{$D^\ast[011] D^\ast[011]\times 2$} &  \textcolor{lightgray}{$D[012] D^\ast[001]\times 2$} \\
        & \textcolor{lightgray}{$D[002] D^\ast[001]$} &  \textcolor{lightgray}{$D[012] D^\ast[001]\times 2$} &  \textcolor{lightgray}{$D^\ast[002] D^\ast[000]$} \\
        & \textcolor{lightgray}{$D^\ast[011] D^\ast[001]\times 4$} &  \textcolor{lightgray}{$D[111] D^\ast[111]\times 2$} &  \textcolor{lightgray}{$D^\ast[011] D^\ast[011]\times 2$} \\
        & \textcolor{lightgray}{$D[011] D^\ast[012]\times 2$} &         & \textcolor{lightgray}{$D[011] D^\ast[112]\times 2$} \\
& &        & \textcolor{lightgray}{$D[112] D^\ast[011]\times 2$} \\

\multicolumn{4}{c}{} \\
\multicolumn{2}{c}{$[011]A_2$} & \multicolumn{2}{c}{$[111]A_2$} \\
\hline
$L/a_s=20$ & $L/a_s=24$ & $L/a_s=20$ & $L/a_s=24$ \\
\hline
$D[000] D^\ast[011]$ & $D[000] D^\ast[011]$ & $D[000] D^\ast[111]$ & $D[000] D^\ast[111]$ \\
$D[001] D^\ast[001]\times 2$ & $D[001] D^\ast[001]\times 2$ & $D[001] D^\ast[011]\times 2$ & $D[001] D^\ast[011]\times 2$ \\
$D[011] D^\ast[000]$ & $D[011] D^\ast[000]$ & $D[111] D^\ast[000]$ & $D[011] D^\ast[001]\times 2$ \\
$D^\ast[011] D^\ast[000]\times 2$ & $D[001] D^\ast[111]\times 2$ & $D[011] D^\ast[001]\times 2$ & $D[111] D^\ast[000]$ \\
 $D^\ast[001] D^\ast[001]\times 2$ & $D[011] D^\ast[011]\times 3$ & $D^\ast[111] D^\ast[000]$ &  $D^\ast[111] D^\ast[000]$\\
 $D[001] D^\ast[111]\times 2$ & $D[111] D^\ast[001]\times 2$ & $D^\ast[011] D^\ast[001]\times 4$ & $D^\ast[011] D^\ast[001]\times 4$ \\
$D[011] D^\ast[011]\times 3$ & $D^\ast[011] D^\ast[000]\times 2$ &  & \textcolor{lightgray}{$D[001] D^\ast[112]\times 2$} \\
$D[111] D^\ast[001]\times 2$ & $D^\ast[001] D^\ast[001]\times 2$ &        & \textcolor{lightgray}{$D[011] D^\ast[012]\times 3$} \\
 & \textcolor{lightgray}{$D[001] D^\ast[012]\times 2$} &         & \textcolor{lightgray}{$D[111] D^\ast[002]\times 2$}\\
        & \textcolor{lightgray}{$D[011] D^\ast[002]\times 2$} &         & \textcolor{lightgray}{$D[112] D^\ast[001]\times 2$} \\
        & \textcolor{lightgray}{$D[012] D^\ast[001]\times 2$} &         & \textcolor{lightgray}{$D[002] D^\ast[111]\times 2$} \\
        & \textcolor{lightgray}{$D[002] D^\ast[011]\times 2$} &         & \textcolor{lightgray}{$D[012] D^\ast[011]\times 3$} \\
        & \textcolor{lightgray}{$D^\ast[111] D^\ast[001]\times 4$} &         &  \\
        & \textcolor{lightgray}{$D^\ast[011] D^\ast[011]\times 3$} &         &  \\
\end{tabular}
\end{small}
\caption{\label{tab:inflight_ops}Operators for the in flight irreps.}
\end{table*}

\clearpage
\pagebreak

\section{\label{sec:cc_hpw}Coupled-Channel Parameterization with $\DstDst$ \SLJ{5}{P}{3}, \SLJ{3}{D}{2,3} and \SLJ{5}{F}{3} Partial Waves}

The reference parameterization of Eq.~(\ref{eq:ref_param}) and the 13 parameterization variations consider $\DstDst$ in \SLJ{3}{S}{1}, \SLJ{3}{D}{1}, \SLJ{5}{P}{2} and \SLJ{5}{F}{2} partial waves. The \SLJ{5}{F}{2} partial wave was introduced in part to enable the description of a set of four levels in the $[111]A_2$ irrep that would be degenerate in the absence of meson-meson interactions. Other $\DstDst$ combinations such as \SLJ{3}{D}{3}, \SLJ{5}{P}{3} and \SLJ{5}{F}{3} also contribute to the $[111]A_2$ irrep and so could potentially be used instead of \SLJ{5}{F}{2}, but the \SLJ{3}{D}{2} partial wave does not contribute to $[111]A_2$ and so could not. To examine the effects of these partial waves, three separate additional variations of the reference parameterization were considered wherein \SLJ{3}{D}{3}, \SLJ{5}{P}{3} and \SLJ{5}{F}{3} were in turn used in place of \SLJ{5}{F}{2}.

\begin{table*}[!h]
\begin{tabular}{|c|c|c|}
\hline
&&\\[-2.5ex]
	$\DstDst$ Partial Wave & $\gamma^{(0)}_{\DstDst \SLJc{2S+1}{\ell}{J} \rightarrow \DstDst \SLJc{2S+1}{\ell}{J}}$ & $\chi^2/N_{\mathrm{dof}}$ \\
&&\\[-2.5ex]
\hline
&&\\[-2.5ex]
\SLJ{3}{D}{3} & $(37 \pm 239) \cdot a_t^4$ & $361.06/(109-14) = 3.80$ \\
\SLJ{5}{P}{3} & $(-9.49 \pm 7.22) \cdot a_t$ & $121.54/(109-14) = 1.28$ \\
\SLJ{5}{F}{3} & $(10800 \pm 48300) \cdot a_t^6$ & $122.08/(109-14) = 1.29$ \\
\hline
\end{tabular}\\
	\caption{\label{tab:pw_exclude}The values of the parameters describing the \SLJ{3}{D}{3}, \SLJ{5}{P}{3} and \SLJ{5}{F}{3} $\DstDst$ scattering amplitudes along with the associated $\chi^2/N_{\mathrm{dof}}$ of the fit.}
\end{table*}

Table \ref{tab:pw_exclude} displays the $\chi^2/N_{\mathrm{dof}}$ and the value of the parameter describing the \SLJ{3}{D}{3}, \SLJ{5}{P}{3} and \SLJ{5}{F}{3} $\DstDst$ scattering amplitudes from these three fits. A poor $\chi^2/N_{\mathrm{dof}}$ is obtained for the parameterization using the \SLJ{3}{D}{3} partial wave because \SLJ{3}{D}{3} does not provide a solution to the determinant condition for reasonable values of the parameter in the energy region of the quadruple degeneracy in the $[111]A_2$ irrep. Additional partial waves are required in order to saturate the observed approximate degeneracy. In contrast, the use of \SLJ{5}{P}{3} or \SLJ{5}{F}{3} yields a $\chi^2/N_{\mathrm{dof}}$ similar to that of the reference parameterization, but does not offer any further improvement. In addition, the parameters describing the \SLJ{5}{P}{3} and \SLJ{5}{F}{3} scattering amplitudes are approximately consistent with zero. This indicates that the finite volume spectrum is insensitive to the choice between \SLJ{5}{P}{3}, \SLJ{5}{F}{2} and \SLJ{5}{F}{3}, and any of these three partial waves can be employed.

We also considered a variation of the reference parameterization where the effects of the $\DstDst$ \SLJ{5}{P}{3}, \SLJ{3}{D}{2,3} and \SLJ{5}{F}{2,3} partial waves were included simultaneously, which accounts for all $\DDst$ and $\DstDst$ partial waves with $\ell \leq 2$ that contribute to the irreps utilized.
Using the reference parameterisation with the addition of a constant term for each of the \SLJ{3}{D}{2,3}, \SLJ{5}{P}{3} and \SLJ{5}{F}{3} $\DstDst$ amplitudes gives,

\begin{small}
\begin{center}
\renewcommand{\arraystretch}{1.4}
\begin{tabular}{rll}
$\gamma^{(0)}_{\DDst \SLJc{3}{S}{1} \rightarrow \DDst \SLJc{3}{S}{1}} = $ & $(6.72 \pm 0.56)$ & \multirow{7}{*}{ $\begin{bmatrix*}[r]   1.00 &  -0.80 &   0.35 &  -0.09 &   0.24 &   0.10 &   0.06\\
&  1.00 &   0.17 &   0.21 &  -0.26 &  -0.01 &  -0.03\\
&&  1.00 &   0.29 &  -0.05 &   0.07 &   0.03\\
&&&  1.00 &  -0.95 &   0.06 &   0.02\\
&&&&  1.00 &  -0.03 &  -0.01\\
&&&&&  1.00 &   0.02\\
&&&&&&  1.00\end{bmatrix*}$ } \\
$\gamma^{(1)}_{\DDst \SLJc{3}{S}{1} \rightarrow \DDst \SLJc{3}{S}{1}} = $ & $(-60 \pm 13) \cdot a_t^2$ & \\
$\gamma^{(0)}_{\DDst \SLJc{3}{S}{1} \rightarrow \DstDst \SLJc{3}{S}{1}} = $ & $(4.16 \pm 0.53)$ & \\
$\gamma^{(0)}_{\DstDst \SLJc{3}{S}{1} \rightarrow \DstDst \SLJc{3}{S}{1}} = $ & $(13.1 \pm 2.6)$ & \\
$\gamma^{(1)}_{\DstDst \SLJc{3}{S}{1} \rightarrow \DstDst \SLJc{3}{S}{1}} = $ & $(-128 \pm 54) \cdot a_t^2$ & \\
$\gamma^{(0)}_{\DDst \SLJc{3}{D}{1} \rightarrow \DDst \SLJc{3}{D}{1}}= $ & $(151 \pm 57) \cdot a_t^4$ & \\
$\gamma^{(0)}_{\DstDst \SLJc{3}{D}{1} \rightarrow \DstDst \SLJc{3}{D}{1}} = $ & $(-831 \pm 267) \cdot a_t^4$ & \\
\end{tabular}
\end{center}

\begin{center}
\renewcommand{\arraystretch}{1.4}
\begin{tabular}{rll}
$\gamma^{(0)}_{\DDst \SLJc{3}{P}{0} \rightarrow \DDst \SLJc{3}{P}{0}} = $ & $(66 \pm 5) \cdot a_t^2$ & \multirow{7}{*}{ $\begin{bmatrix*}[r]   1.00 &   0.06 &   0.12 &   0.00 &   0.26 &  -0.02 &   0.00\\
&  1.00 &   0.13 &   0.00 &  -0.70 &  -0.03 &   0.00\\
&&  1.00 &   0.00 &  -0.03 &  -0.75 &   0.00\\
&&&  0.00 &   0.00 &   0.00 &   0.00\\
&&&&  1.00 &   0.00 &   0.00\\
&&&&&  1.00 &   0.00\\
&&&&&&  0.00\end{bmatrix*}$ } \\
$\gamma^{(0)}_{\DDst \SLJc{3}{P}{2} \rightarrow \DDst \SLJc{3}{P}{2}} = $ & $(30 \pm 4) \cdot a_t^2$ & \\
$\gamma^{(0)}_{\DstDst \SLJc{5}{P}{2} \rightarrow \DstDst \SLJc{5}{P}{2}} = $ & $(77 \pm 18) \cdot a_t^2$ & \\
$\gamma^{(0)}_{\DstDst \SLJc{5}{P}{3} \rightarrow \DstDst \SLJc{5}{P}{3}} = $ & $(-5.80 \pm 21.0) \cdot a_t^2$ & \\
$\gamma^{(0)}_{\DDst \SLJc{3}{F}{2} \rightarrow \DDst \SLJc{3}{F}{2}} = $ & $(-391 \pm 998) \cdot a_t^6$ & \\
$\gamma^{(0)}_{\DstDst \SLJc{5}{F}{2} \rightarrow \DstDst \SLJc{5}{F}{2}} = $ & $(-1680 \pm 56840) \cdot a_t^6$ & \\
$\gamma^{(0)}_{\DstDst \SLJc{5}{F}{3} \rightarrow \DstDst \SLJc{5}{F}{3}} = $ & $(18233 \pm 70489) \cdot a_t^6$ & \\
\end{tabular}
\end{center}

\begin{center}
\renewcommand{\arraystretch}{1.4}
\begin{tabular}{rll}
$\gamma^{(0)}_{\DDst \SLJc{3}{D}{2} \rightarrow \DDst \SLJc{3}{D}{2}} = $ & $(82 \pm 64)  \cdot a_t^4$ & \multirow{4}{*}{ $\begin{bmatrix*}[r]   1.00 &   0.05 &  -0.02 &   0.03\\
&  1.00 &   0.03 &   0.04\\
&&  1.00 &  -0.14\\
&&&  1.00\end{bmatrix*}$ } \\
$\gamma^{(0)}_{\DDst \SLJc{3}{D}{3} \rightarrow \DDst \SLJc{3}{D}{3}} = $ & $(-78 \pm 31)  \cdot a_t^4$ & \\
$\gamma^{(0)}_{\DstDst \SLJc{3}{D}{2} \rightarrow \DstDst \SLJc{3}{D}{2}} = $ & $(-3229 \pm 1134)  \cdot a_t^4$ & \\
$\gamma^{(0)}_{\DstDst \SLJc{3}{D}{3} \rightarrow \DstDst \SLJc{3}{D}{3}} = $ & $(352 \pm 262)  \cdot a_t^4$ & \\[1.3ex]
&\multicolumn{2}{l}{ $\chi^2/ N_\mathrm{dof} = \frac{116.4}{109-18} = 1.28$\,.}
\end{tabular}
\end{center}
\end{small}
The $\chi^2 / N_{\mathrm{dof}}$ is the same as in the reference parameterization, suggesting that the inclusion of these additional partial waves is not necessary to describe the finite volume spectrum. Furthermore, the values of parameters describing the $J^P = 1^+$ scattering amplitudes display very little deviation between this parameterization and the reference parameterization. The central values of these parameters in this parameterization are well contained within the statistical uncertainty of the parameters in the reference parameterization. It is also observed that the statistical uncertainties of the parameters common to this and the reference parameterization display very little deviation between the two parameterizations. This indicates that these partial waves have little effect on the scattering amplitudes in $J^P = 1^+$ and can be neglected, justifying the approach of Sec. \ref{sec:coupled}.

\clearpage
\pagebreak

\section{\label{sec:svdlims}SVD Limits of the Data Correlation Matrix}
As discussed in Section \ref{sec:coupled}, we employ the use of singular value \textit{limits} in the correlated $\chi^2$ minimization used for the parameter determination. It would be preferable to perform the $\chi^2$ minimization without any manipulation of the underlying correlations of the data, however large data correlations between energy levels on each of the three lattice volumes result in an unreasonable large $\chi^2/N_\mathrm{dof}$.   

Large correlations between data points can arise when the smallest eigenvalues of the data correlation matrix are underestimated. This can occur when the number of Monte Carlo samples is of the order of the square of the number of data points, which is precisely the situation in this work. Across all irreps and lattice volumes, we have utilized 109 energy levels. The largest $L/a_s=24$ volume provided 56 energy levels to constrain the scattering amplitudes while having $553$ gauge configurations (Monte Carlo samples). It is then not possible to reliably determine the data covariance matrix as the number of independent numbers in the correlation matrix is greater than the number of Monte Carlo samples. 

In singular value limits, the singular values (eigenvalues) of the symmetric correlation matrix undergo the replacement $\lambda_i \rightarrow \mathrm{max}(\lambda_i, \sigma)$ for $\sigma = \tau \lambda_1$, where $\tau$ is the singular value limit. This is in contrast to singular value \emph{cuts}, where the modes are removed from the data covariance matrix inverse.

In Appendix D of Ref. \cite{Dowdall:2019bea}, singular value limits are shown to be equivalent to adding an additional statistical uncertainty to the data (see Eq.~D17 therein), as the replacement sets the eigenvalue to at least its true value. As such, taking a singular value limit can be seen as a conservative approach. Singular value cuts can be interpreted in the same way, however the replacement sets the eigenvalue to $\infty$, thus making it the most extreme and conservative case of singular value limits. This can be ameliorated by modifying the interpretation of the $\chi^2/N_\mathrm{dof}$ as a measure of the quality of the fit. When a singular value cut is used, it is to natural reduce the number of degrees of freedom by the number of modes cut, such that $\hat{N}_{\mathrm{dof}}= N_\mathrm{data}-N_\mathrm{cut}-N_\mathrm{pars}$, as was done in Refs.~\cite{Cheung:2020mql, Wilson:2023anv}. For singular value limits, we retain the usual definition, $N_{\mathrm{dof}}= N_\mathrm{data}-N_\mathrm{pars}$.

\begin{table*}[!h]
\begin{tabular}{|c|c|c|}
\hline
&&\\[-2.5ex]
SVD limit    & Number of Replaced Modes   &  \multirow{2}{*}{$\chi^2/N_\mathrm{dof}$} \\
$\tau$       & (total: 6+47+56)           &                                           \\
\hline
0                      & $0 +  0 +  0$   & $759.10/(109-14) \quad = \quad 7.99$ \\
0.003                  & $0 + 26 + 33$   & $284.28/(109-14) \quad = \quad 2.99$ \\
0.006                  & $0 + 34 + 40$   & $177.24/(109-14) \quad = \quad 1.87$ \\ 
$\!\!\!^{\dag}\bm{0.010}$   & $\bm{0 + 37 + 46}$   & $\bm{121.26/(109-14) =\;\; 1.28}$ \\
$0.015$                & $0 + 40 + 48$   &  $\;\: 88.39/(109-14) \quad = \quad 0.93$ \\
$0.020$                & $0 + 42 + 49$   &  $\;\: 69.35/(109-14) \quad = \quad 0.73$ \\
\hline
\end{tabular}\\
\begin{tabular}{|c|c|c|}
\hline
&&\\[-2.5ex]
$\;\;$ SVD cut $\;\;$   & $\;\;\:$ Number of Modes Cut $\;\;\:$ &  \multirow{2}{*}{$\chi^2/\hat{N}_\mathrm{dof}$} \\
$\tau$     & (total: 6+47+56)          &         \\
\hline
0.0016 & $0 + 21 + 23$       & ${119.65}/(109-14-44)$ = 2.35\\
0.0020 & $0 + 22 + 27$       & ${69.60}/(109-14-49)$  = 1.52\\
0.0024 & $0 + 25 + 29$       & ${62.01}/(109-14-54)$  = 1.51\\
0.0032 & $0 + 28 + 33$       & ${48.90}/(109-14-61)$  = 1.44\\
0.0040 & $0 + 31 + 37$       & ${35.48}/(109-14-68)$  = 1.31\\ 
\hline
\end{tabular}\\
        \begin{tabular}{|c|c|c|}
                \hline
                &&\\[-2.5ex]
		   & $\;\;\:$ Total Number of Modes $\;\;\:$ &  $\chi^2/N_\mathrm{dof}$ \\
		\hline
		uncorrelated & $6+47+56$ & $169.51/(109-14) \quad = \quad 1.78$ \\
                \hline
        \end{tabular}
	\caption{\label{tab:svd_lim_tab}The number of eigenmodes replaced, $\chi^2$ and $\chi^2 / N_\mathrm{dof}$ for each value of the singular value limit/cut $\tau$. The number of modes replaced for the three lattice volumes is denoted as $(16^3 + 20^3 + 24^3)$. The value of $\tau$ used in the main text is denoted with $\dag$  in bold. }
\end{table*}

Performing a range of amplitude determinations with various choices of SVD limits and cuts, we find a range of reasonable choices.
Table~\ref{tab:svd_lim_tab} shows the value of the $\chi^2$ for the reference parameterization (Eq.~(\ref{eq:ref_param})) for fully correlated ($\tau = 0$) data and five values of $\tau$, including the number of eigenvalues replaced for each lattice volume. In Table~\ref{tab:svd_lim_tab}, we also show the $\chi^2$ values obtained when using SVD cuts, as well as the $\chi^2$ for uncorrelated data. For any of the fits listed in Table~\ref{tab:svd_lim_tab} with a $\chi^2/N_\mathrm{dof}\lesssim 2$, the central values of the determined amplitudes and poles are largely unchanged within uncertainties. In general, larger values of $\tau$ lead to broader uncertainties.

\begin{figure*}[!h]
\includegraphics{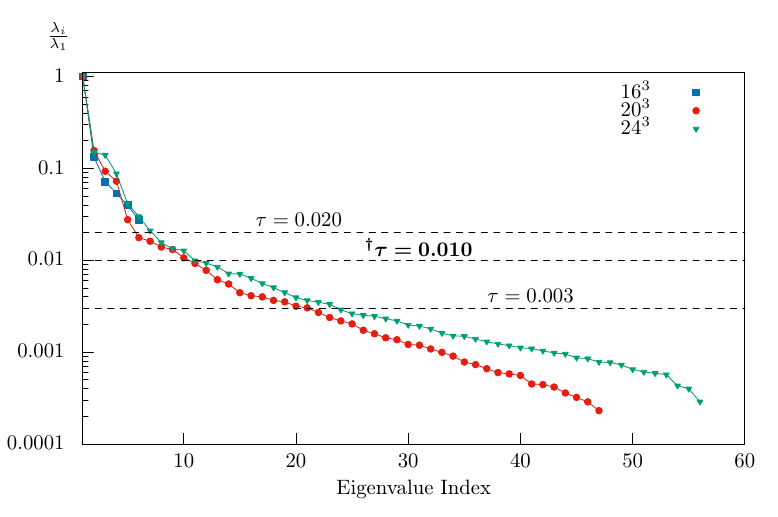}
\caption{\label{fig:svd} The eigenvalues of the correlation matrix on each lattice volume normalized by the largest eigenvalue $\lambda_1$. The value of $\tau$ used in the main analysis is denoted with $\dag$ in bold.}
\end{figure*}

Fig. \ref{fig:svd} shows the eigenvalues of the data correlation matrix for each lattice volume, normalized by the largest eigenvalue $\lambda_1$. It is observed that the $\chi^2$ for the fully correlated parameterization is large and cannot be trusted as a valid parameterization. Using a singular value limit of $\tau = 0.003$ results in a decrease of the $\chi^2$, but not enough to have confidence in the parameterization. A limit of $\tau = 0.01$ results in a reasonable $\chi^2/N_\mathrm{dof}$. The $\chi^2$ is lower than that of the uncorrelated fit, which reflects the additional uncertainty introduced by this approach. 

\clearpage
\twocolumngrid
\bibliography{refs.bib}
\bibliographystyle{apsrev}

\end{document}